\documentclass[aip,pop,reprint,groupedaddress, floatfix]{revtex4-1}

\usepackage{graphicx}
\newcommand{\be}{\begin{displaymath}}
\newcommand{\bn}{\begin{equation}}

\newcommand{\en}{\end{equation}}
\newcommand{\ee}{\end{displaymath}}

\usepackage{times}
\usepackage{amssymb}
\usepackage{amsmath}
\usepackage{url,hyperref}

\graphicspath{{Bilderfertig/inFile/}}

\begin{document}
\title{Collisionless microinstabilities in stellarators II - numerical simulations}
\author{J.~H.~E.~Proll} \author{P.~Xanthopoulos} \author{P.~Helander}
\affiliation{Max-Planck-Institut f\"ur Plasmaphysik, EURATOM Association,
Teilinstitut Greifswald, Wendelsteinstra{\ss}e 1, 17491 Greifswald, Germany}

\begin{abstract}
Microinstabilities exhibit a rich variety of behavior in stellarators due to
the many degrees of freedom in the magnetic geometry. It has recently been
found that certain stellarators (quasi-isodynamic ones with maximum-$J$
geometry) are partly resilient to trapped-particle instabilities, because
fast-bouncing particles tend to extract energy from these modes near
marginal stability.
In reality, stellarators are never perfectly quasi-isodynamic, and the
question thus arises whether they still benefit from enhanced stability.
Here the stability properties of Wendelstein 7-X and a more quasi-isodynamic
configuration, QIPC, are investigated numerically and compared with the
National Compact Stellarator Experiment (NCSX) and the DIII-D tokamak. In
gyrokinetic simulations, performed with the gyrokinetic code GENE in the
electrostatic and collisionless approximation, ion-temperature-gradient
modes, trapped-electron modes and mixed-type instabilities are studied.
Wendelstein 7-X and QIPC exhibit significantly reduced growth rates for all
simulations that include kinetic electrons, and the latter are indeed found
to be stabilizing in the energy budget. 
These results suggest that imperfectly optimized stellarators can retain
most of the stabilizing properties predicted for perfect maximum-$J$
configurations.
\end{abstract}

\maketitle
\normalsize
\section{Introduction}
The effect of three-dimensional magnetic fields on the confinement properties of fusion devices is of great interest. With the increasing awareness that tokamaks are never perfectly axisymmetric, because of error fields or intentionally applied resonant magnetic perturbations, and the development of new stellarator concepts, this question deserves renewed attention. 
The three-dimensional geometry of various stellarators can have very different character, and presumably so can the stability properties. Traditionally, stellarators suffered from large neoclassical transport, while in tokamaks the turbulent transport usually dominates. Optimized stellarator configurations, such as the National Compact Stellarator Experiment (NCSX) \cite{Zarnstorff2001} and Wendelstein 7-X (W7-X)\cite{Beidler1990}, or more recently proposed geometries, such as the quasi-isodynamic configuration of Subbotin et al.~\cite{Subbotin2006}, whose contours of constant magnetic field strength are poloidally closed and we therefore refer to as QIPC, are designed to have reduced neoclassical transport \cite{Beidler2011}. The question then arises whether this improvement will lead to reduced turbulent transport caused by microinstabilities such as the trapped-electron mode (TEM) or ion temperature gradient driven modes (ITGs). Such a correlation between reduced neoclassical and turbulent transport has been suggested by Watanabe et al.~ \cite{Watanabe2008} as a consequence of reduced zonal-flow damping, but we are more concerned with the direct effect of the neoclassical optimization on the linear microinstabilities. 

In Part I of the present publication, this question was addressed analytically, with the result that perfectly quasi-isodynamic configurations were found to be largely resilient against collisionless TEMs as well as the collisionless trapped-particle mode, if the parallel adiabatic invariant $J$ decreases with radius, the so-called maximum-$J$ property. In such configurations, the electron diamagnetic drift frequency $\omega_{*e}$ has the opposite sign from the bounce-averaged magnetic drift frequency of the electrons, $\overline{\omega}_{de}$ for all orbits on the flux surface, i.e. $\overline{\omega}_{de}\omega_{*e}<0$. The electrons were then predicted to extract energy from each mode near marginal stability rather than to pump energy into the mode. Since quasi-isodynamicity and maximum-$J$ geometry cannot be achieved exactly, it is interesting to assess to which extent the criterion $\overline{\omega}_{de}\omega_{*e}<0$ must be fulfilled in order to still retain the stabilizing effect of the kinetic electrons. Instabilities including kinetic electrons have not been studied extensively in stellarators. Although there are some results focusing on W7-X \cite{Xanthopoulos2007,Xanthopoulos2007b} and NCSX \cite{Baumgaertel2011, Baumgaertel2012b, Baumgaertel2013} as well as comparisons of different stellarators \cite{Rewoldt2005}, these are not sufficient to explain why certain stellarators are more resilient against microinstabilities than others. In the present paper, two stellarators approaching quasi-isodynamicity, W7-X and QIPC, are compared with the DIII-D tokamak and the NCSX stellarator, which are both distinctly non-quasi-isodynamic. By varying the temperature and density gradients, and by making simulations with both adiabatic and kinetic electrons, the role of the latter is elucidated.
\\
In section \ref{sec:geometries}, the different configurations that will be simulated are introduced and analyzed with respect to regions of bad curvature and as to whether (and where) the stability criterion $\overline{\omega}_{de}\omega_{*e}<0$ is met. After briefly introducing the simulation setup in section \ref{sec:GENE}, we study the various instabilities in section \ref{sec:results}, ranging from ITGs with adiabatic electrons over ITGs with kinetic electrons to classical TEMs and mixed ITG-TEM modes. Here the focus lies on analyzing the structure of the different types of modes that are found and comparing their growth rates. In section \ref{sec:conclusions}, conclusions are drawn and an outlook to further work is provided.
\section{The geometries}
\label{sec:geometries}
Exploring the influence of the geometry of a configuration on the stability of microinstabilities is the main goal of this paper. We therefore devote the present section to introducing the different geometries used - starting with an axisymmetric equilibrium corresponding to the DIII-D tokamak, followed by the quasi-axisymmetric NCSX stellarator and ending with two more quasi-isodynamic configurations, W7-X and QIPC, an almost perfectly quasi-isodynamic stellarator (particularly at high plasma pressure)\cite{Subbotin2006}.\\
The main feature of interest is how well the stability criterion identified in part I of this publication, $\overline{\omega}_{de}\omega_{*e}<0$, is met in the respective configurations, since we then expect a stabilizing influence on the TEM. For ITGs, on the other hand, it is not the bounce-averaged curvature but rather the local curvature that drives the instability. The electron drift frequency in a plasma with zero pressure can be expressed as 
$$\omega_{de}=\mathbf{k}_{\bot}\cdot \hat{\mathbf{b}}\times \nabla_{\bot}\ln B \left(\frac{v^2_{\bot}}{2}+v^2_{\|}\right)/\Omega_e,$$
where $\Omega_e=-eB/m_e$ is the electron cyclotron frequency and $\mathbf{k}_{\bot}$ the perpendicular wave vector. We use the the GIST geometry interface \cite{Xanthopoulos2009} to calculate the geometry-dependent part of this expression, $\kappa \propto \left(\hat{\mathbf{b}}\times \nabla_{\bot}\ln B\right) \cdot \mathbf{k}_{\bot}$ and display the result following the GENE convention  that $\kappa<0$ indicates a ``bad-curvature'' region, i.e. where the ITG modes are expected to be located. \\
For all configurations that are studied, a flux surface with normalized toroidal flux of $s=\psi/\psi_0=0.25$ was chosen, which corresponds to the surface at half radius. The magnetic field structure for the stellarator cases was calculated using the VMEC code \cite{Hirshman1986} assuming zero beta, to be consistent with the neglect of electromagnetic effects in the gyrokinetic simulations. These were carried out in the flux-tube approximation, using the two flux tubes on the surface in question that are stellarator symmetric. A more detailed description of the numerical aspects can be found in Ref.~\cite{Xanthopoulos2007b}. For our first configuration, the DIII-D tokamak, all flux tubes are of course equivalent.
\subsection{The DIII-D tokamak}
\begin{figure}
\includegraphics[width=0.45 \textwidth]{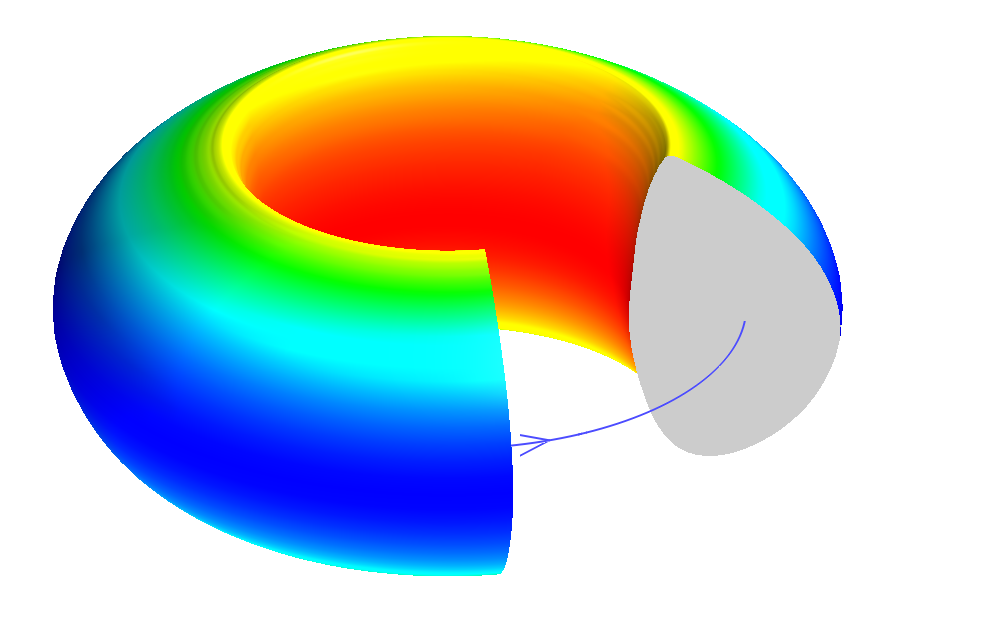}
\caption{\label{fig:d3dvacB3d}Magnetic field strength $B$ of DIII-D, red representing the maximum of the field, blue the minimum.}
\end{figure}

\begin{figure}
\includegraphics[width=0.5 \textwidth]{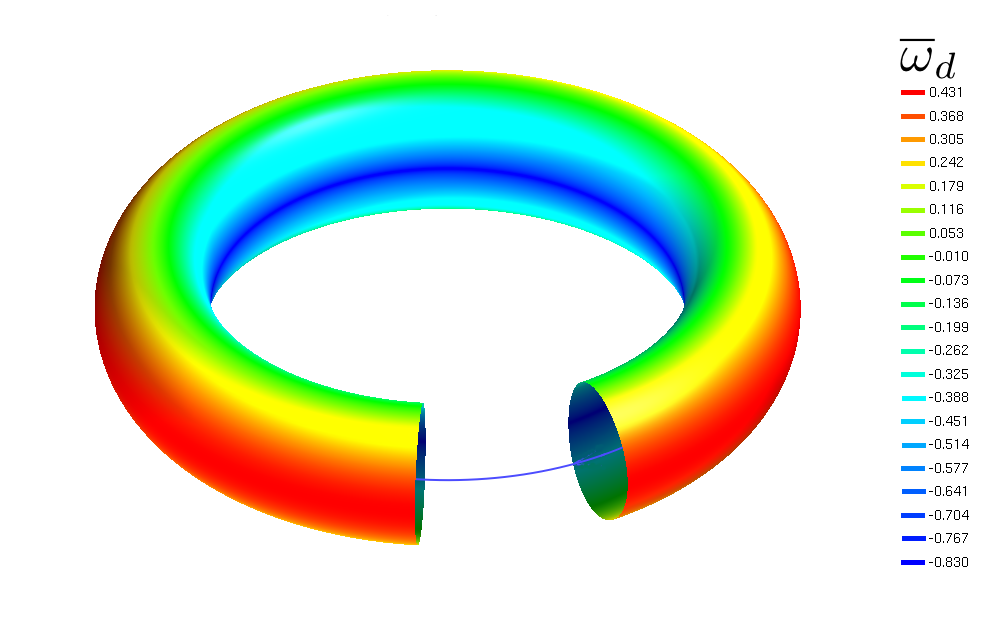}
\caption{\label{fig:d3dvacomegad}Drift frequency $\overline{\omega}_{de}$ on the flux surface at half radius of DIII-D, as a function of bounce point location.}
\end{figure}

\begin{figure}
\includegraphics[width=0.45 \textwidth]{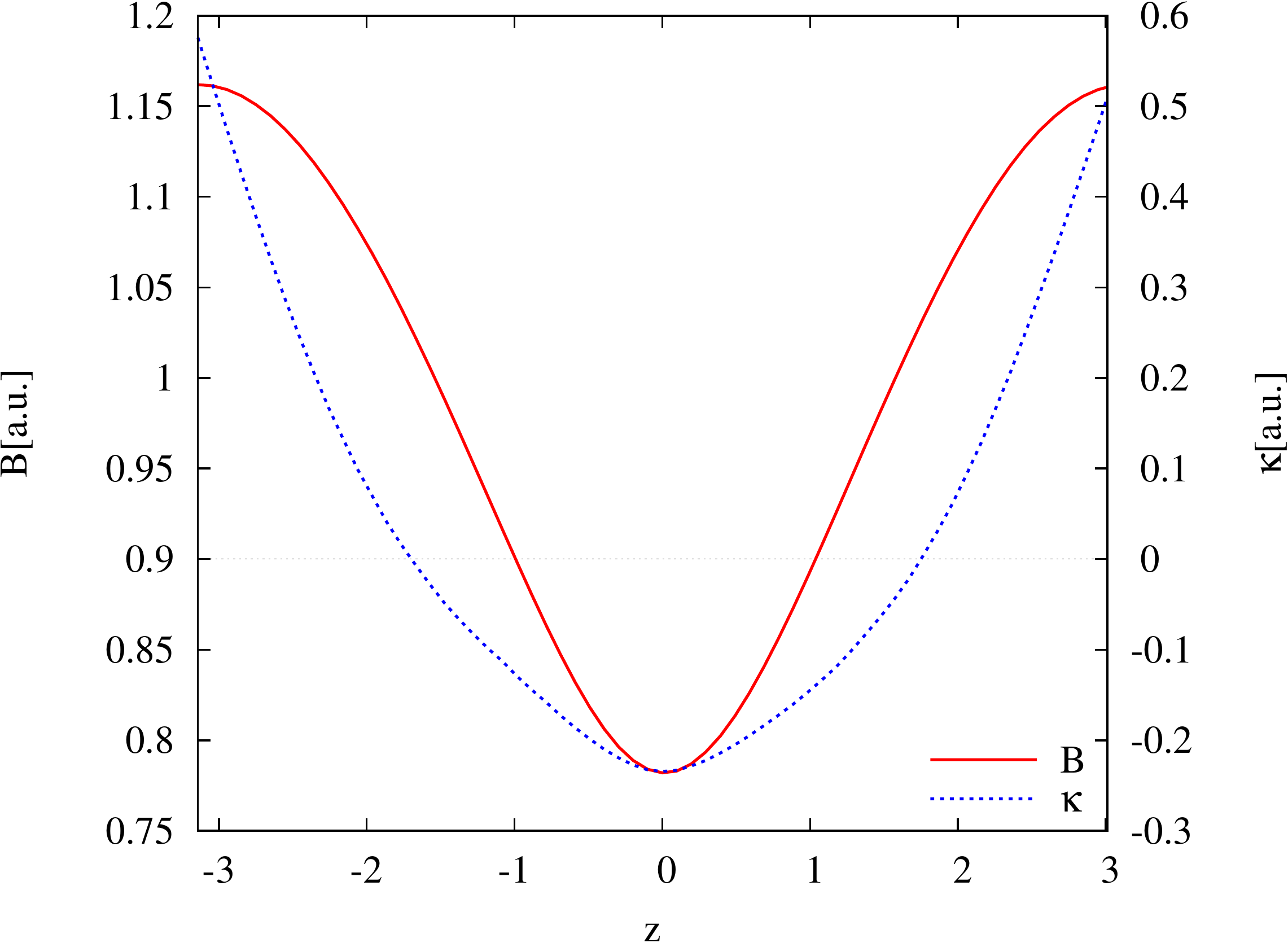}
\caption{\label{fig:geometryd3dvac} Magnetic field strength $B$ and curvature $\kappa$ along the magnetic field line, with $z=0$ in the outboard midplane}
\end{figure}

The DIII-D tokamak was chosen in order to compare the stellarator data with a typical tokamak configuration. The EFIT file used for creating the flux tube stems from the DIII-D discharge $\#$128913, which has been the subject of previous gyrokinetic studies \cite{Bravenec2011}. As can be seen in Fig.\ \ref{fig:d3dvacB3d}, the poloidal cut has a D-shaped cross-section.
The sign of the wave vector is chosen so that $\omega_{*e}$ is positive, and as seen from Fig.\ \ref{fig:d3dvacomegad}, the precessional drift $\overline{\omega}_{de}$ is then also positive for orbits whose orbits lie on the outboard side. The TEM stability criterion is thus violated in the region were most trapped particles are located, and a significant level of TEM activity can thus be expected. This is typical of tokamaks as can be seen from the following equation, which gives the toroidal precession in a tokamak depending on the bounce parameter $k$, which is $k>1$ for passing particles and $0<k<1$ for trapped particles\cite{Kadomtsev1970, HelanderSigmar}
\begin{equation}
\overline{\dot{\phi}}=\frac{q^2 v^2}{\Omega Rr}\left[\frac{E(k)}{K(k)}-\frac{1}{2}+\frac{2rq'}{q}\left(\frac{E(k)}{K(k)}+k^2-1\right)\right].
\end{equation}
Here, $q$ is the safety factor and $q'$ the magnetic shear, $\Omega=e_a B/m$ denotes the gyro frequency, and $R$ and $r$ are the major and the minor radius, respectively. $K(k)$ and $E(k)$ are the complete elliptic integrals of the first and second kind. The terms independent of the magnetic shear (the two first terms in the brackets) show that, because $E(k)/K(k)$ varies between $1$ for deeply trapped particles with $k=0$ and $0$ for barely trapped particles with $k=1$, these differently trapped particles precess in opposite directions. The deeply trapped particles obviously reside in a region of bad curvature, whilst the barely trapped particles spend most of their time on the inboard side of the torus and precess in the opposite toroidal direction. If $q'=0$ the precession of deeply trapped particles therefore tends to be resonant with the diamagnetic frequency whilst shallowly trapped particles are not resonant.
The term proportional to the magnetic shear has a similar effect if the shear is negative in the tokamak sense, $q'<0$. Deeply trapped particles then have an additional positive contribution from $q'<0$ to the precession drift, while the contribution is negative for barely trapped particles. If, however, the magnetic shear is positive, as is the case for this particular DIII-D configuration, the sign of the contribution proportional to the shear is reversed.
The local curvature is depicted in Fig.\ \ref{fig:geometryd3dvac}, showing a clear overlap of the magnetic well with the region of negative $\kappa$ (``bad curvature'');  both the magnitude of the magnetic field $B$ and the curvature $\kappa$ have their minimum at the outboard midplane. 
This overlap of the magnetic well, which is mainly important for TEMs, and the ``bad curvature'', usually associated with ITGs, can make it difficult to discriminate between TEMs and ITGs when looking at the mode structure alone.
\subsection{NCSX}
\begin{figure}
\includegraphics[width=0.45 \textwidth]{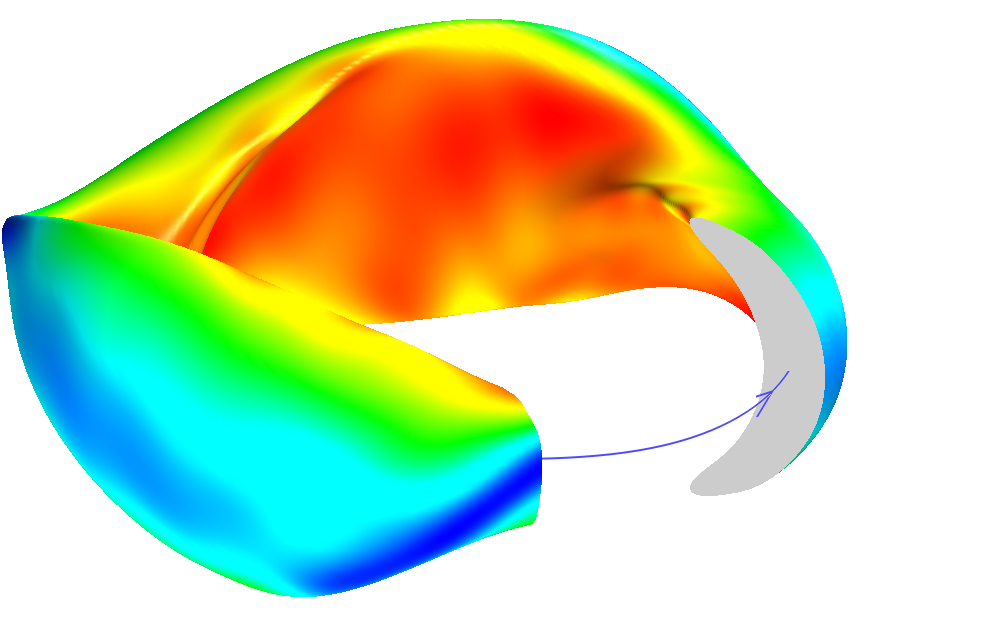}
\caption{\label{fig:ncsxvacbeanB3d}Magnetic field strength $B$ of NCSX, red representing the maximum of the field, blue the minimum. The poloidal cut shows the so-called bean plane.}
\end{figure}

\begin{figure}
\includegraphics[width=0.45 \textwidth]{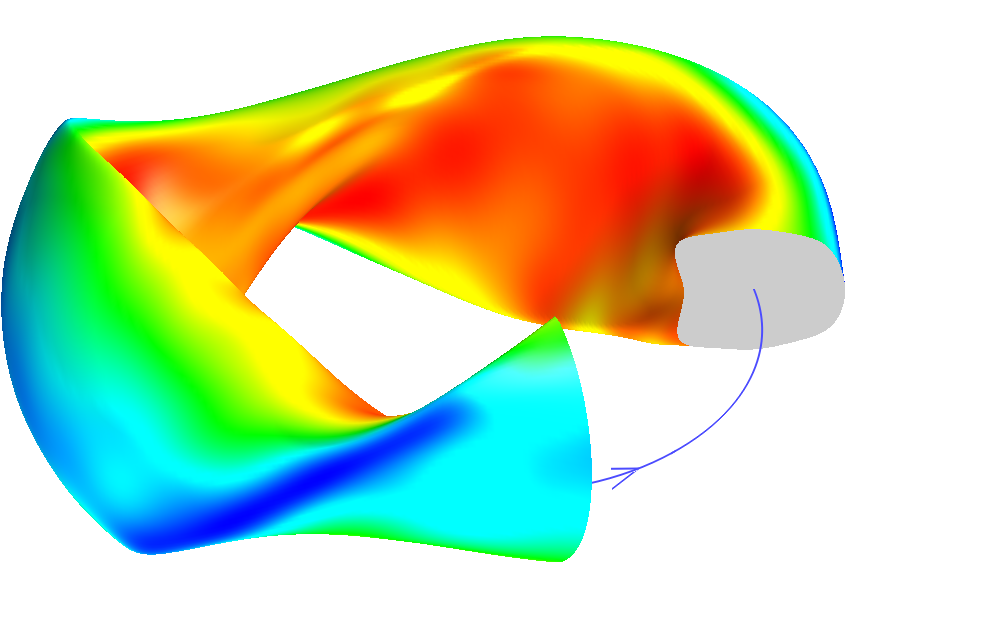}
\caption{\label{fig:ncsxvacbulletB3d}Magnetic field strength $B$ in NCSX, red representing the maximum of the field, blue the minimum. The poloidal cut shows the so-called bullet plane.}
\end{figure}

\begin{figure}
\includegraphics[width=0.5 \textwidth]{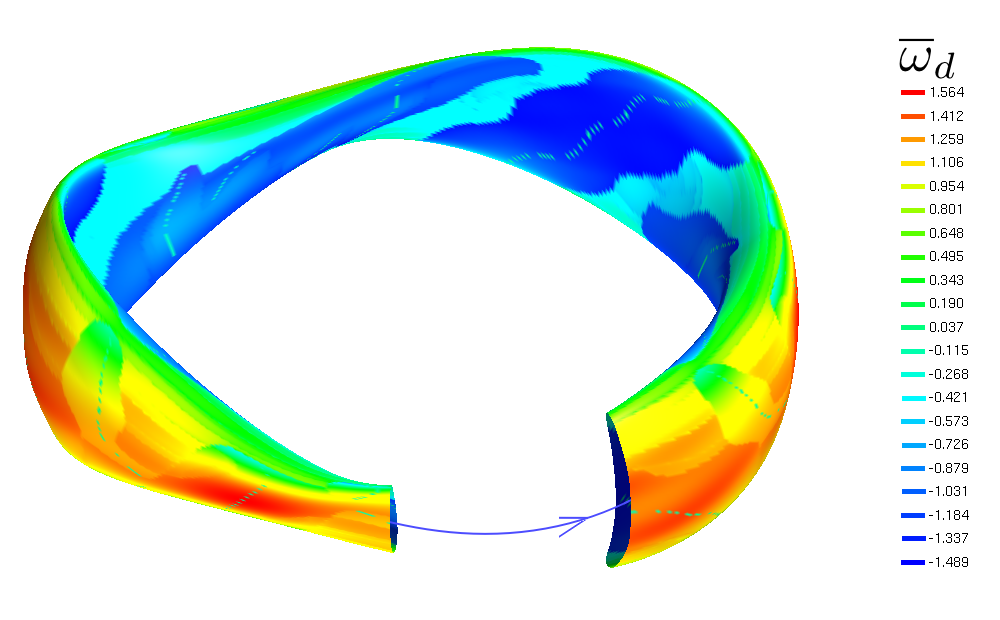}
\caption{\label{fig:ncsxvacomegad}$\overline{\omega}_{de}$ on the flux surface at half radius of NCSX, as a function of bounce point location.}
\end{figure}

\begin{figure}
\includegraphics[width=0.45 \textwidth]{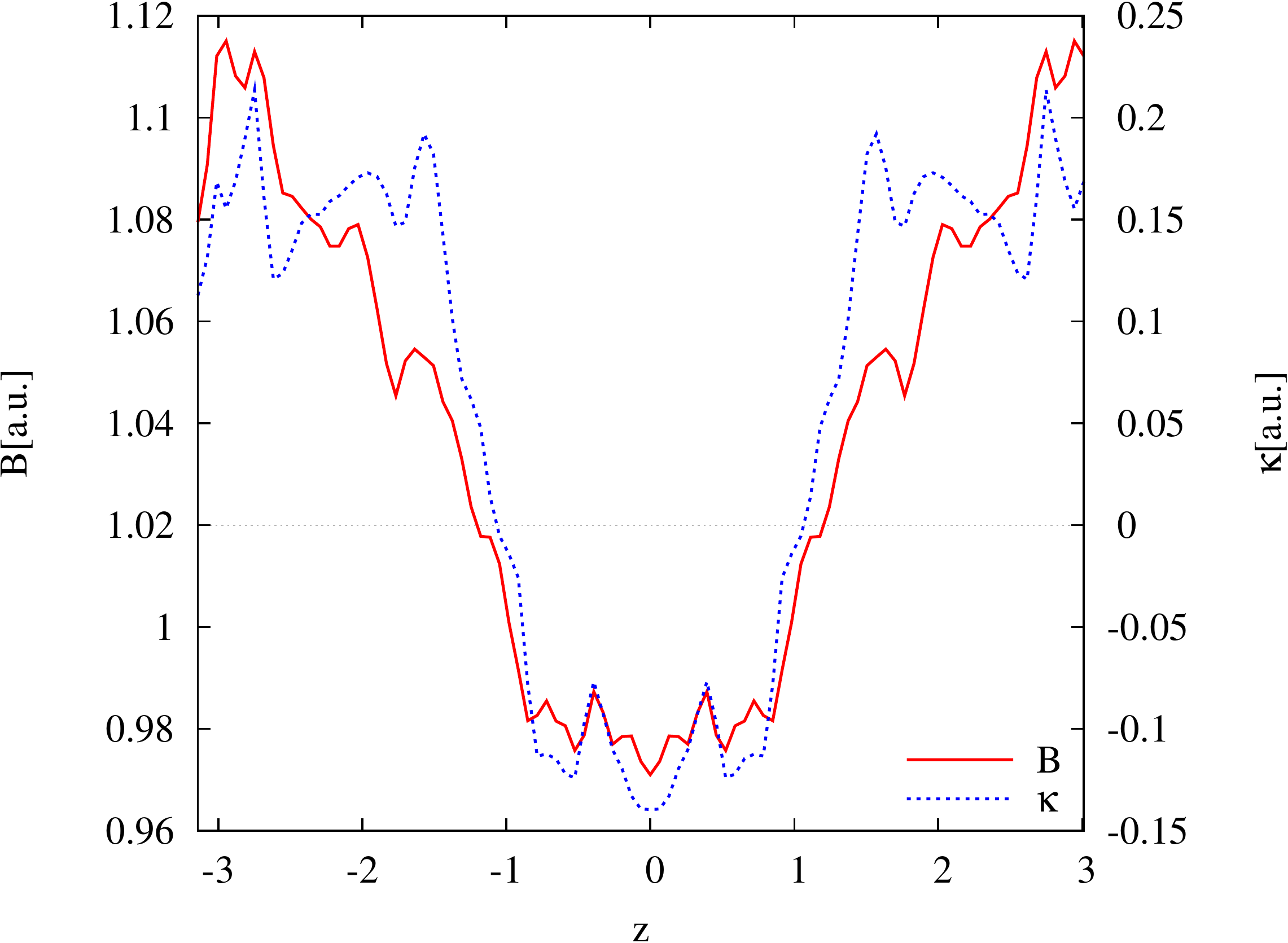}
\caption{\label{fig:geometryncsxvacbean} Magnetic field strength $B$ and curvature $\kappa$ along a magnetic field line in NCSX. $z = 0$ in the outboard midplane of the bean plane.}
\end{figure}

\begin{figure}
\includegraphics[width=0.45 \textwidth]{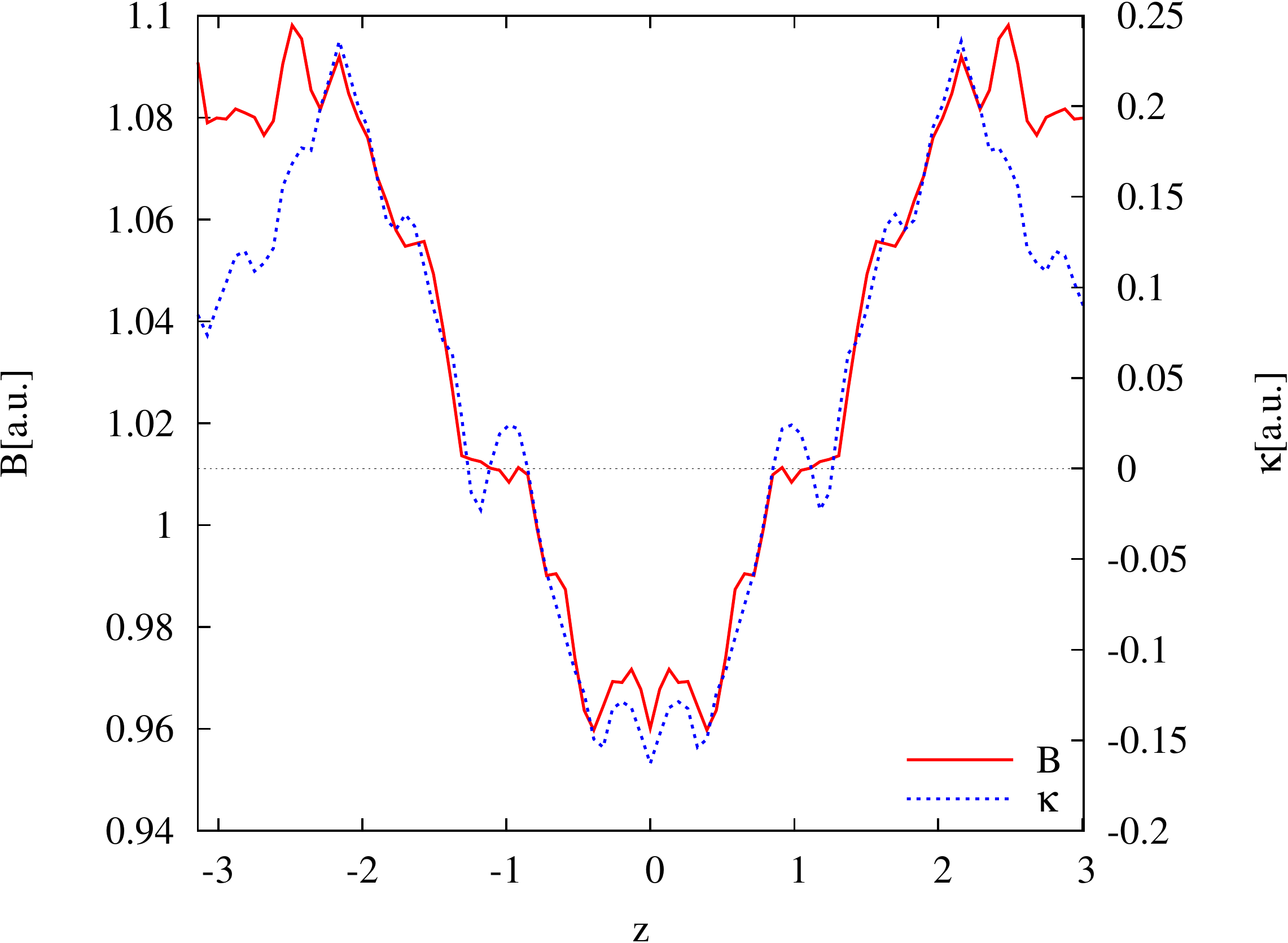}
\caption{\label{fig:geometryncsxvacbullet} Magnetic field strength $B$ and curvature $\kappa$ along the magnetic field line centered around the outboard midplane of the bullet plane of NCSX.}
\end{figure}

It is enlightening to have a different kind of stellarator to compare the results of the nearly quasi-isodynamic stellarators with. NCSX was chosen since it provides an intermediate step between a tokamak and, for example, W7-X when it comes to aspect ratio and three-dimensionality. In Fig.\ \ref{fig:ncsxvacbeanB3d} and \ref{fig:ncsxvacbulletB3d}, the magnitude of the magnetic field is displayed. NCSX is a quasi-axisymmetric stellarator \cite{Nuhrenberg1994} with a threefold symmetry. As in DIII-D, the maximum of the field is found on the inboard side and the minimum on the outboard side of the torus. 
Because of the loss of axisymmetry the flux tubes are not equivalent. Therefore two different flux tubes are chosen to investigate stability, one with its center in the outboard midplane of the so-called bean-shaped plane, see Fig.\ \ref{fig:ncsxvacbeanB3d}, the other centered around the outboard midplane of the ``bullet'' plane, which is separated from the bean-shaped plane by a toroidal angle of $\pi/3$ (see Fig.\ \ref{fig:ncsxvacbulletB3d}).
With regard to the bounce-averaged magnetic drift in Fig.\ \ref{fig:ncsxvacomegad}, NCSX behaves similarly to a tokamak - on the outboard side the precessional drift $\overline{\omega}_{de}$ is resonant with the electron drift wave frequency $\omega_{*e}$, so that the TEM stability criterion is only met by orbits bouncing on the inboard side.
The distribution of magnetic field strength $B$ and curvature $\kappa$ also behave as in a tokamak - both have their minimum in the outboard midplane for both flux tubes (Figs.\ \ref{fig:geometryncsxvacbean} and \ref{fig:geometryncsxvacbullet}).
\subsection{Wendelstein 7-X}
\begin{figure}
\includegraphics[width=0.45 \textwidth]{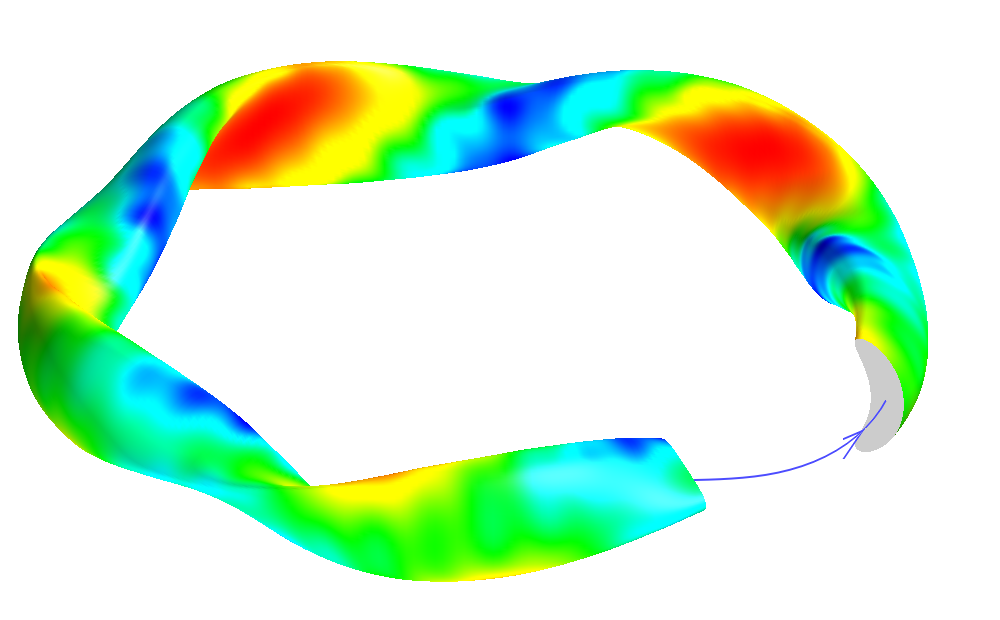}
\caption{\label{fig:w7xvacbeanB3d}Magnetic field B of W7-X, red depicting the maximum of the field, blue the minimum. The visible cut shows the so-called bean plane.}
\end{figure}

\begin{figure}
\includegraphics[width=0.45 \textwidth]{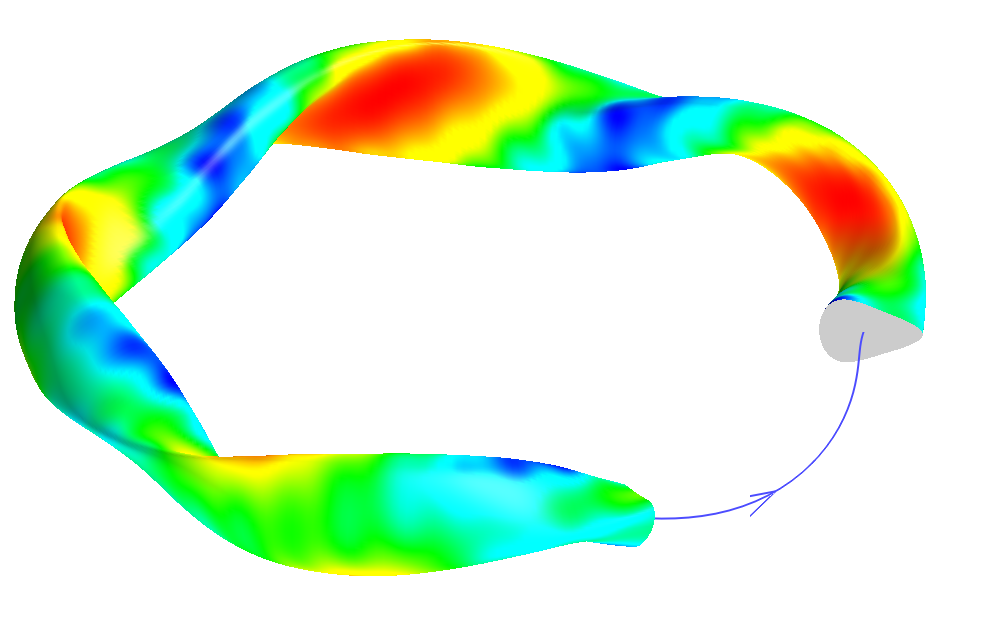}
\caption{\label{fig:w7xvactriangleB3d}Magnetic field strength $B$ in W7-X. The poloidal cut is taken in the triangular plane.}
\end{figure}

\begin{figure}
\includegraphics[width=0.5 \textwidth]{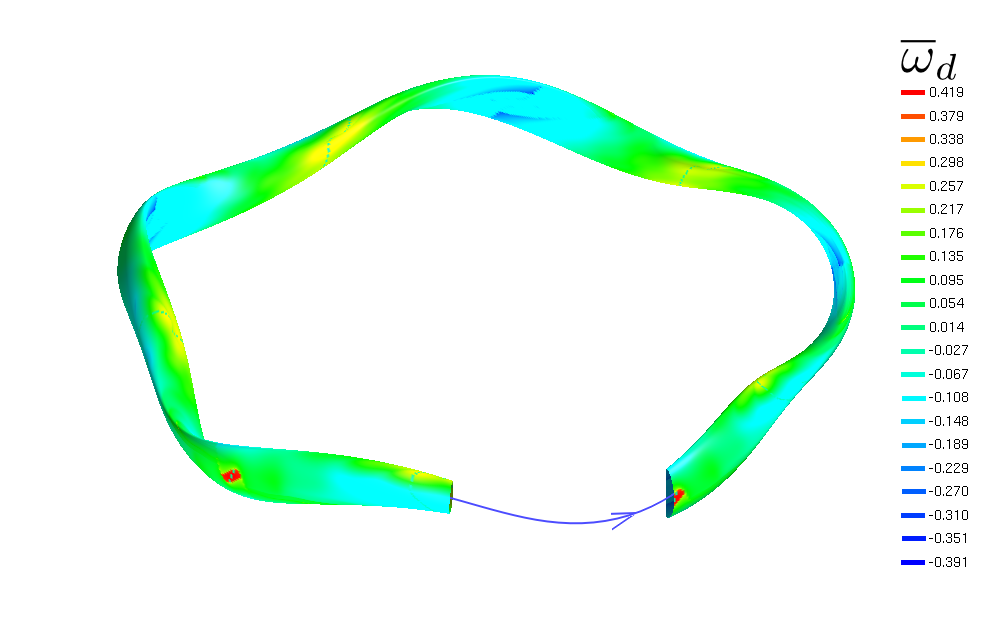}
\caption{\label{fig:w7xvacomegad}$\overline{\omega}_{de}$ on the flux surface at half radius of W7-X, as a function of bounce point location.}
\end{figure}

\begin{figure}
\includegraphics[width=0.45 \textwidth]{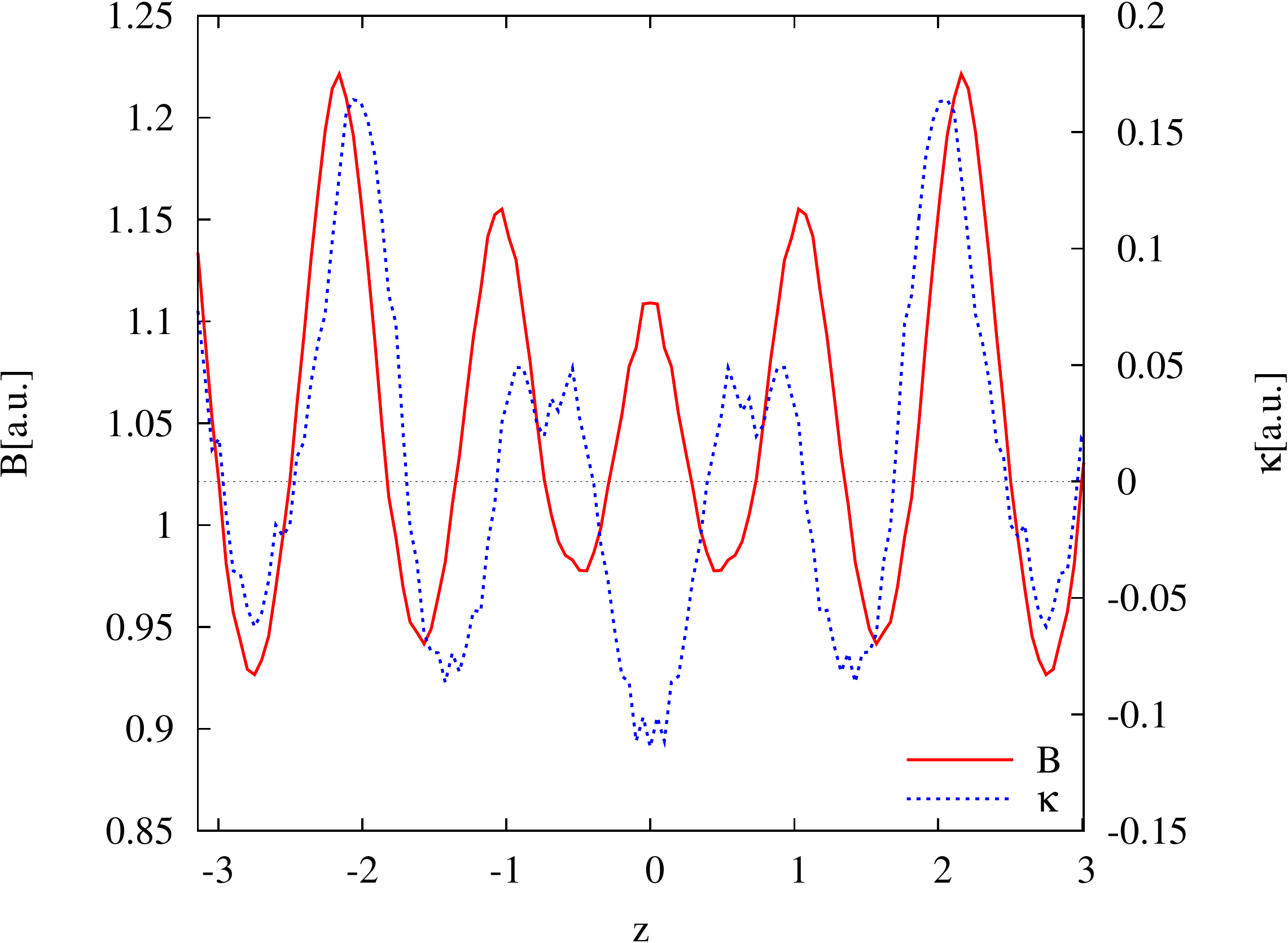}
\caption{\label{fig:geometryw7xvacbean} Magnetic field strength $B$ and curvature $\kappa$ along the magnetic field line along the field line that cuts the outboard midplane (at $z=0$) of the bean-shaped poloidal cross section of W7-X.}
\end{figure}

\begin{figure}
\includegraphics[width=0.45 \textwidth]{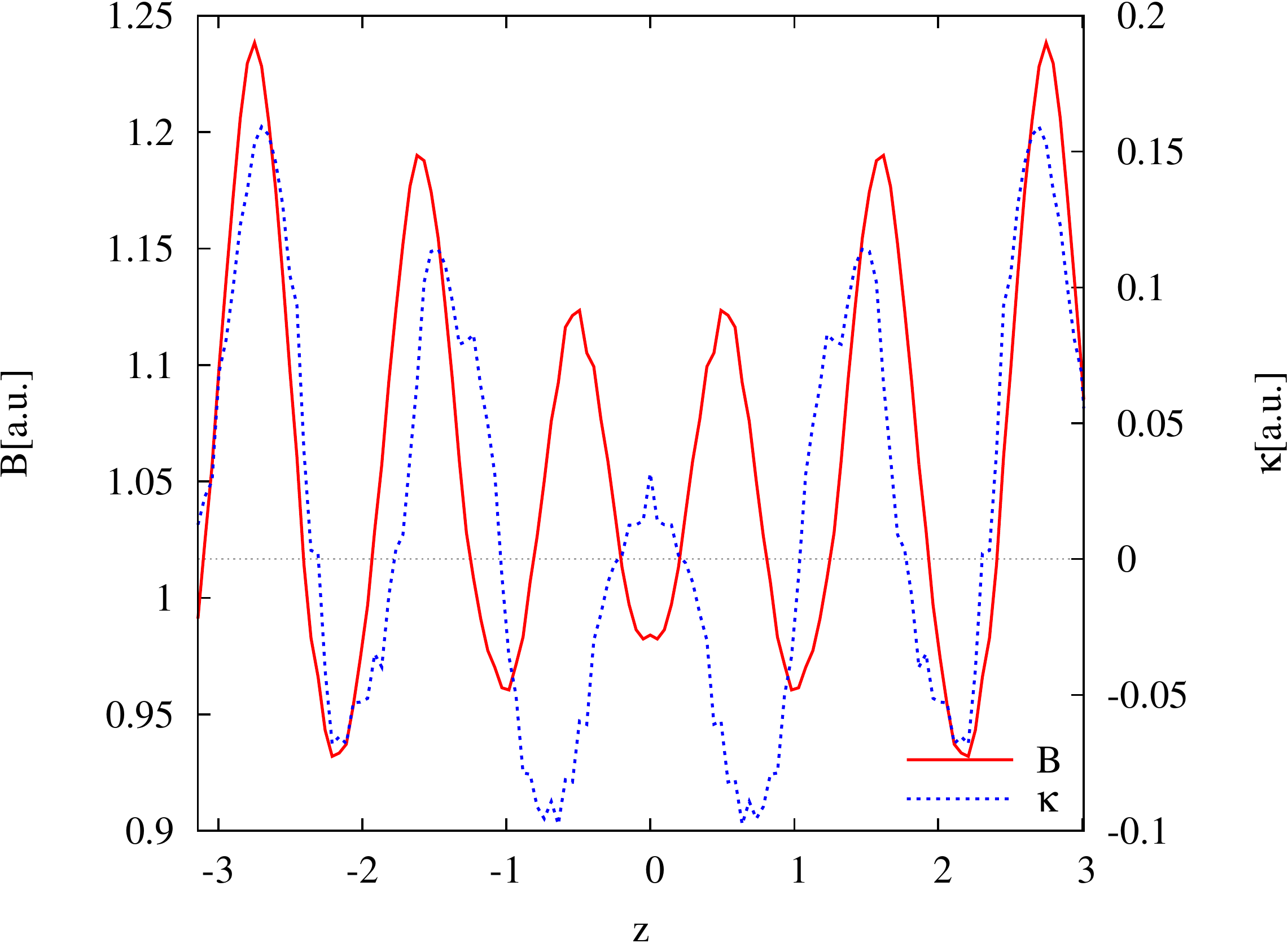}
\caption{\label{fig:geometryw7xvactriangle} Magnetic field strength $B$ and curvature $\kappa$ along the magnetic field line passing through the outboard midplane in the triangular cross section of W7-X.}
\end{figure}

W7-X is the latest stellarator along the Helias line \cite{Grieger1992}, optimized for strongly reduced neoclassical transport \cite{Beidler2011} and approaching quasi-isodynamicity at high beta. It is five-fold symmetric with the maximum of the magnetic field situated in the inner corners of the pentagon and with almost straight (when viewed from above) sections connecting these corners (see Fig.\ \ref{fig:w7xvacbeanB3d}).
As for NCSX, two flux tubes are investigated, one centered in the outboard midplane of the bean-shaped cross-section (Fig.\ \ref{fig:w7xvacbeanB3d}) and the other one in the so-called triangular plane (Fig.\ \ref{fig:w7xvactriangleB3d}), which is $\pi/5$ further along the toroidal direction.
The bounce-averaged drift frequency (Fig.\ \ref{fig:w7xvacomegad}) has negative values in the corners of the device and positive values in the straight sections, where the stability criterion is thus not met. However, the amount of such bad curvature is much lower than in DIII-D and NCSX, and there is, furthermore,
some separation between the regions with trapped particles and bad curvature: the minima of $B$ and the minima of the local curvature $\kappa$ are shifted with respect to each other for both flux tubes used (Figs.\ \ref{fig:geometryw7xvacbean} and \ref{fig:geometryw7xvactriangle}), especially at the center of each flux tube, which will enable us to distinguish between TEMs and ITGs by examining their mode structure.
\subsection{The QIPC stellarator}
\begin{figure}
\includegraphics[width=0.45 \textwidth]{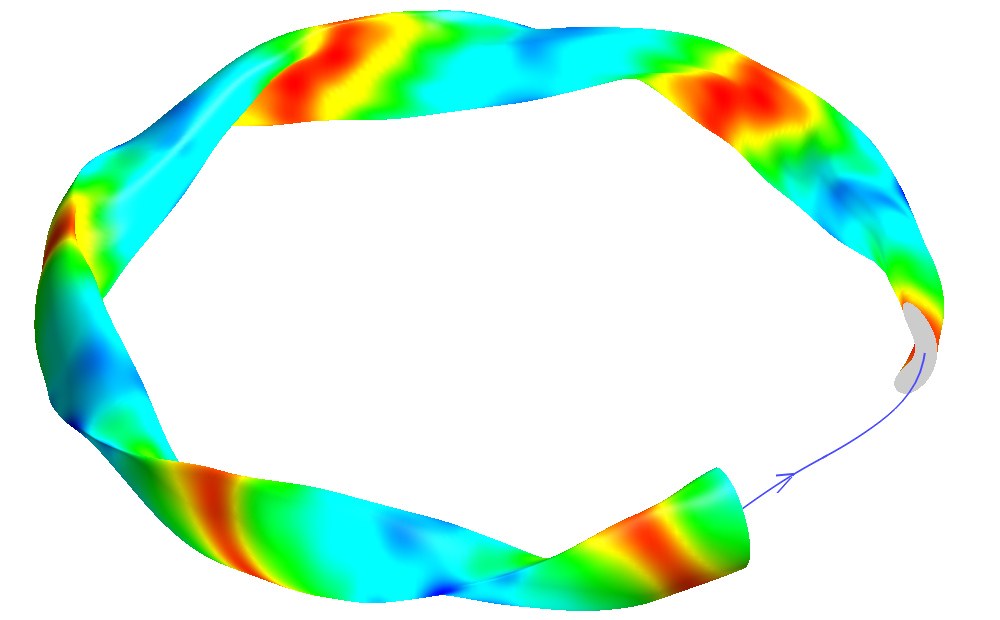}
\caption{\label{fig:qivacbeanB3d}Magnetic field strength $B$ in the QIPC stellarator. The poloidal cut shows the so-called bean plane.}
\end{figure}

\begin{figure}
\includegraphics[width=0.45 \textwidth]{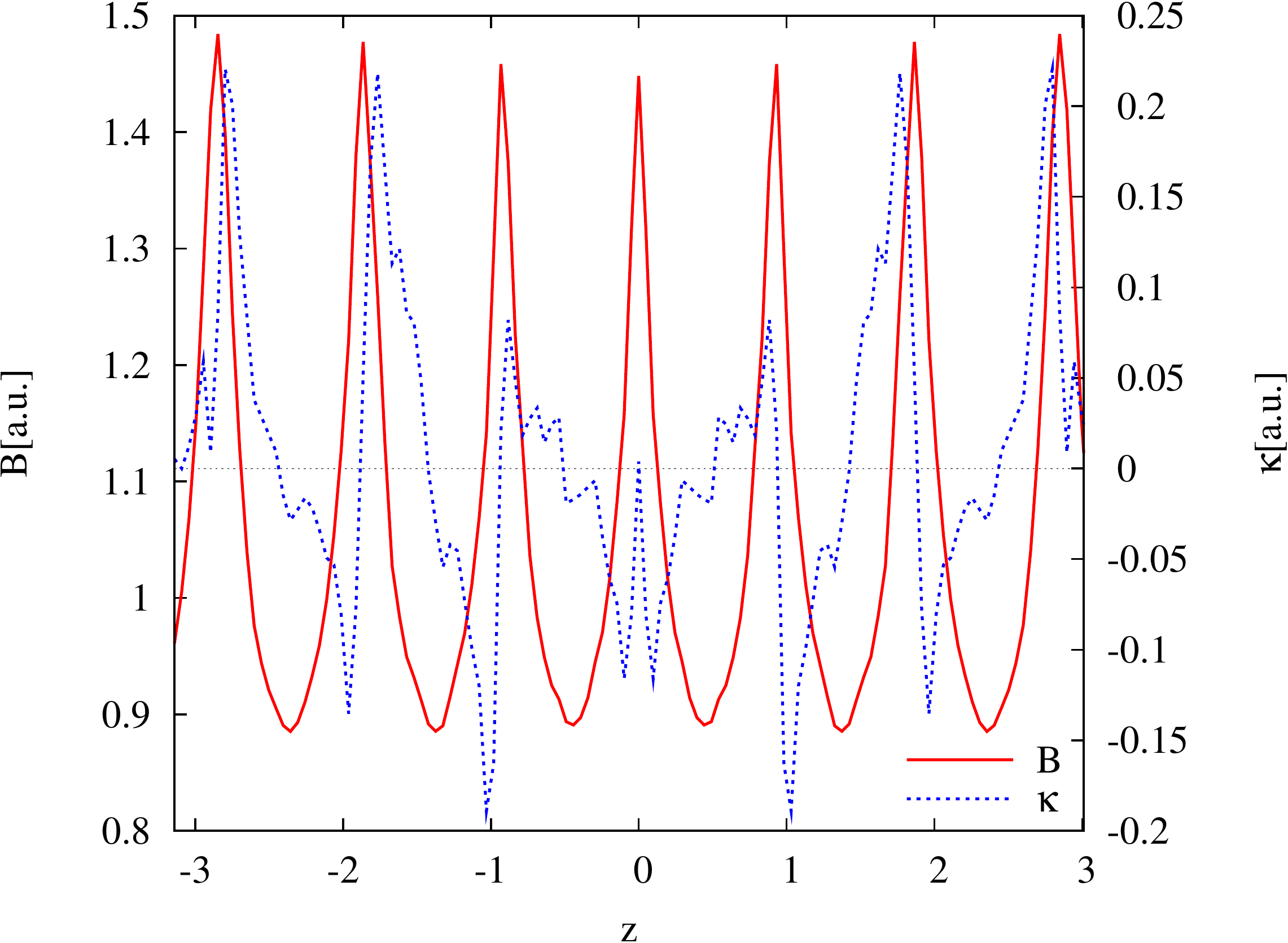}
\caption{\label{fig:geometryqivacbean} Magnetic field strength $B$ and curvature $\kappa$ along the magnetic field line cutting the outboard midplane in the bean-shaped cross section of QIPC.}
\end{figure}

\begin{figure}
\includegraphics[width=0.45 \textwidth]{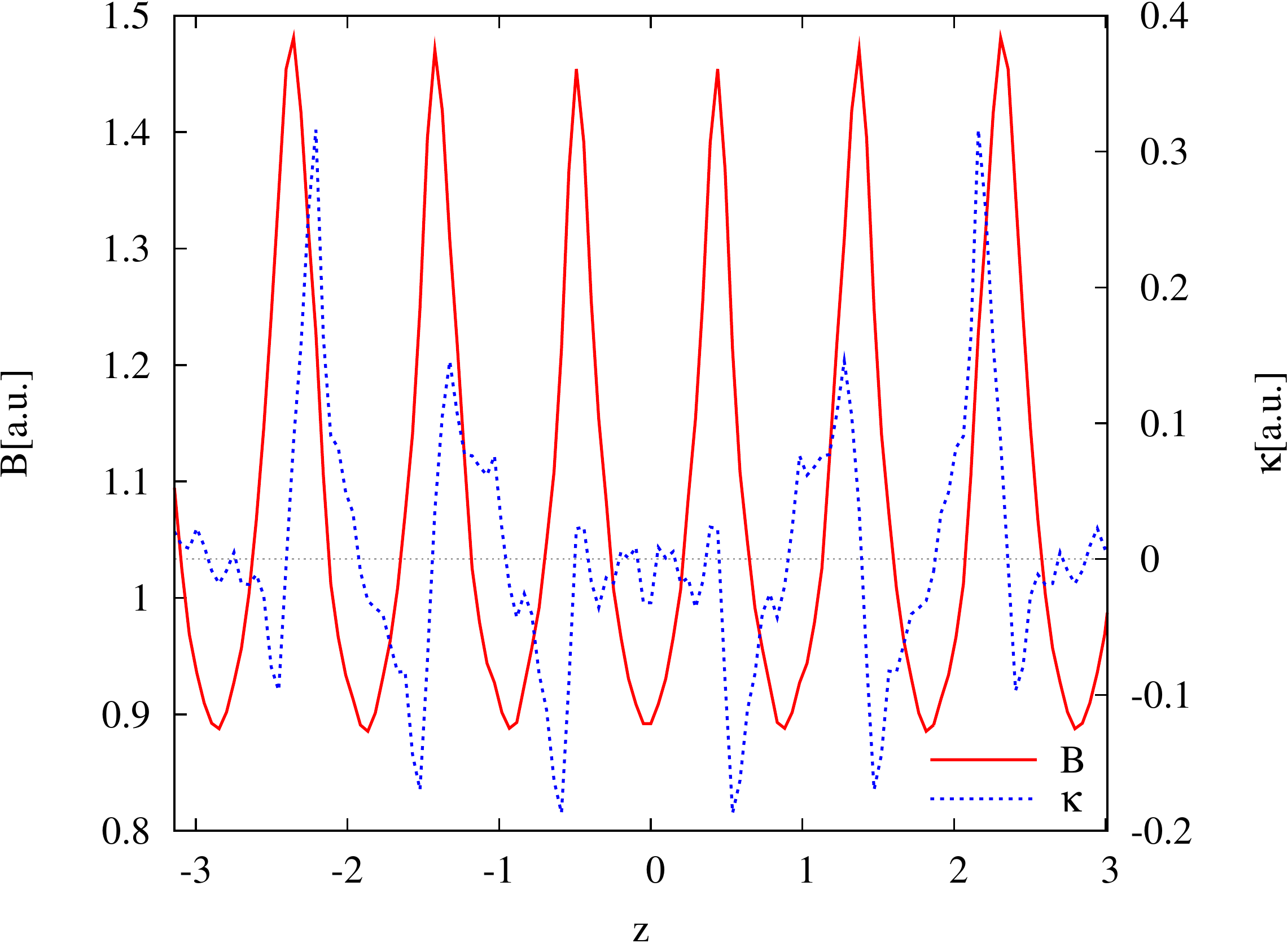}
\caption{\label{fig:geometryqivactriangle} Magnetic field strength $B$ and curvature $\kappa$ along the magnetic field line passing through the outboard midplane in the triangular cross section of QIPC. }
\end{figure}

The QIPC stellarator has the most quasi-isodynamic (``QI'') field of our various configurations. It also belongs to the Helias line and has a sixfold symmetry. As a result of its QI optimization, it has very low neoclassical transport and bootstrap current \cite{Helander2009}. In contrast to W7-X, the contours of constant magnetic field $B$ are indeed poloidally closed (``PC'') (see Fig.\ \ref{fig:qivacbeanB3d}), which can also be seen in Figs. \ref{fig:geometryqivacbean} and \ref{fig:geometryqivactriangle}, where the magnitude of the minima and maxima of $B$ are constant along the field. As in W7-X, the maxima of the field are located in the corners, where the poloidal cross section is bean-shaped, and the minima are located in the straight sections, where the cross section is triangular.
\begin{figure}
\includegraphics[width=0.45 \textwidth]{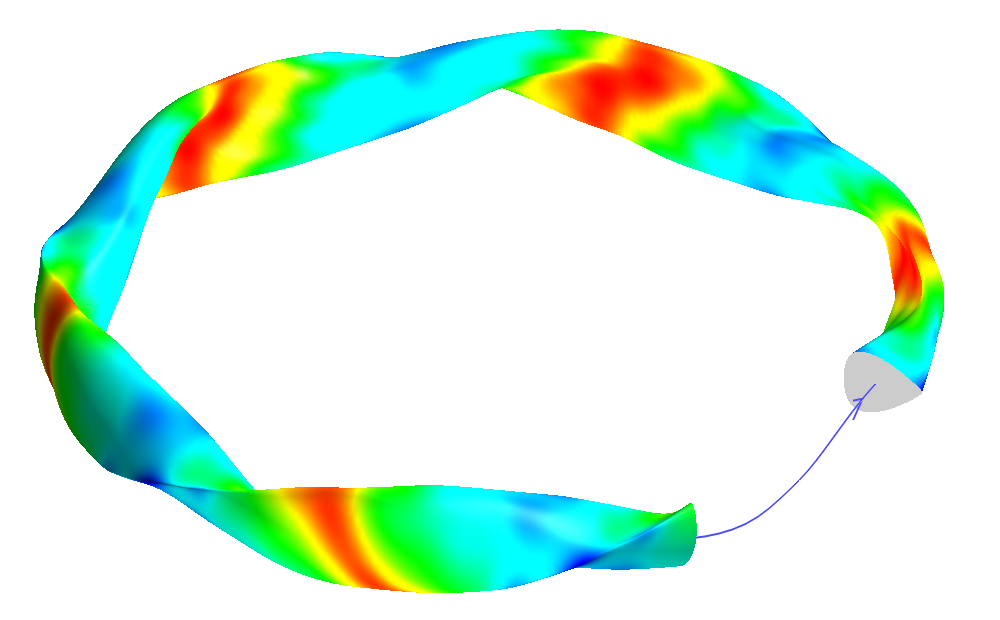}
\caption{\label{fig:qivactriangleB3d}Magnetic field strength $B$ of the QIPC stellarator. The poloidal cut shows the so-called triangular plane.}
\end{figure}
The similarities with W7-X also extend to the distribution of the bounce-averaged magnetic drift frequency. It is in the straight sections where $\overline{\omega}_{de}$ is resonant with the diamagnetic drift $\omega_{*e}>0$ and the stability criterion is mildly violated. The flux surfaces within  $s=0.25$ are, however, almost perfectly quasi-isodynamic when the normalized plasma pressure $\beta$ exceeds a few $\%$.
\begin{figure}
\includegraphics[width=0.5 \textwidth]{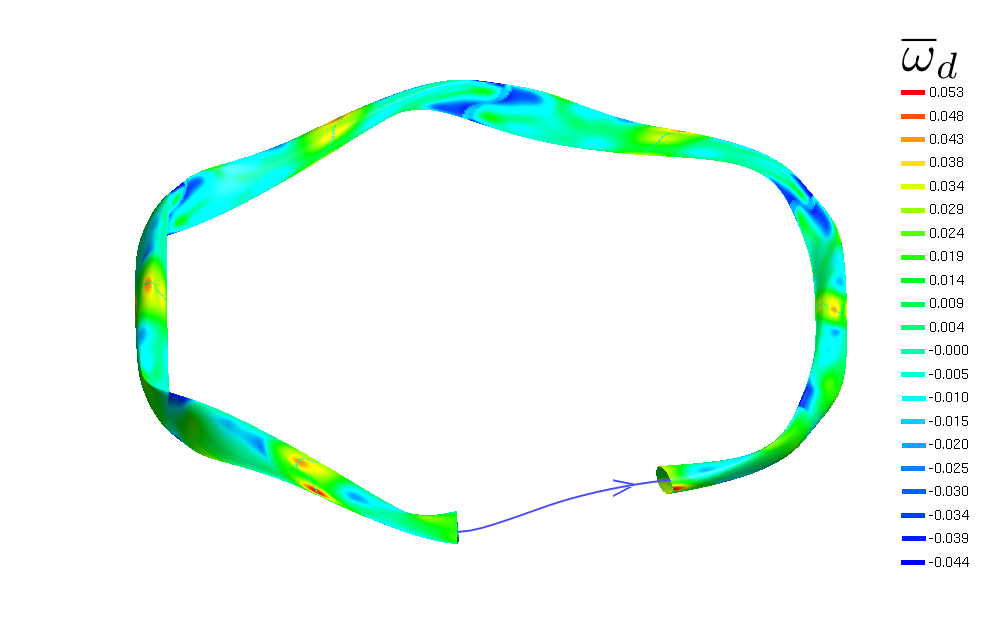}
\caption{\label{fig:qivacomegad}$\overline{\omega}_{de}$ on the flux surface at half radius of the QIPC stellarator, as a function of bounce point location.}
\end{figure}
In QIPC the magnetic wells are slightly more separated from the negative curvature regions than in W7-X. The regions of bad curvature are remarkably small, which can be observed for both flux tubes (Figs.\ \ref{fig:geometryqivacbean} and \ref{fig:geometryqivactriangle}).
Henceforth we will refer to the flux tube centered around the bean-shaped cross section as the ``bean flux tube'', and let the ``triangle flux tube'' and the ``bullet flux tube'' denote the tubes centered around the triangular plane or the bullet-shaped plane in NCSX, respectively.

\section{Gyrokinetic simulation code}
\label{sec:GENE}
The simulations were performed with the gyrokinetic Vlasov code GENE\cite{Jenko2000}, which solves the gyrokinetic equation along with Maxwell's equations. GENE is capable of performing linear and nonlinear simulations, but here only linear instability calculations will be presented. Recently, GENE has been extended to be able to treat the entire plasma radius (excluding a small region around the magnetic axis) in a tokamak \cite{Gorler2011} or an  entire flux surface in a stellarator \cite{Helander2012}. Here, however the domain chosen for the simulations is that of a flux tube, which greatly reduces the computational cost compared with global simulations.\\
In our simulations, VMEC equilibria in case of the stellarators, or the EFIT equilibrium of DIII-D, were transformed into flux-tube geometry data using the GIST geometry interface \cite{Xanthopoulos2009}. When creating this data, the resolution along the field line, in the $z$-direction, must be chosen. From experience, in order to properly resolve the mode structure of trapped-electron modes, each magnetic well needs at least 16 points along the field line. Accordingly, the number of points chosen for DIII-D was $nz0=64$, and because of the less smooth structure of the field in NCSX we chose $nz0=96$. W7-X and QIPC, which have 5 and 6 wells, respectively, were both simulated with $nz0=128$. For the radial direction and the two velocity space coordinates, separate convergence tests were conducted, for each class of instabilities separately in each configuration. To give a typical example, TEMs in the bean flux tube of W7-X required $nkx=8$ radial modes, $nv0=64$ points in the parallel velocity coordinate, and $nw0=16$ points for the magnetic moment coordinate. These values can of course be different for other types of modes and other geometries.\\
In order to avoid unphysical modes, GENE can employ hyperdiffusion \cite{Pueschel2010a}. The hyperdiffusivity in the direction along the field was set to $\epsilon_z=0.5$, and the one in the parallel velocity direction $\epsilon_v=0.2$ in the notation of Ref.\ \cite{Pueschel2010a}. The mass ratio of electrons and hydrogen ions was fully retained, i.e. $m_i/m_e=1836$.\\
After the convergence tests, the range of most unstable modes was determined for each class of instabilities in either flux tube. Thus a scan in the binormal wave number, normalized by the sound gyroradius, $k_y\rho_s$, was performed. The binormal wave number is directly related to the perpendicular wave vector ${\bf k}_{\bot}=k_{\alpha}\nabla \alpha + k_{\psi}\nabla \psi$ where $k_{\alpha}=q k_y a$ with the minor radius $a$ and the safety factor $q$.
If a very detailed analysis is desired, the growth rate could be maximized over both $k_y$ and $k_{\psi}$. However, in tokamaks and also in the $\alpha=0$ flux-tubes of stellarators (that is, where the toroidal angle $\phi$ is zero at the outboard midplane, where the poloidal angle $\theta=0$ as well and therefore $\alpha=\theta-\iota\phi=0$), the modes mainly peak on the outboard side. Setting $k_{\psi}=0$ will therefore give the highest growth rates. 
We do vary $k_{\psi}$ somewhat by choosing two different flux-tubes. This can be seen when writing
\begin{align*}
{\bf k}_{\bot}&= k_{\psi,1}\nabla \psi+ k_{\alpha,1}\nabla \alpha\\
&= k_{\psi,1}\nabla \psi+ k_{\alpha,1} \left(\nabla \theta -\iota \nabla \phi -\iota' \phi\nabla \psi\right),
\end{align*}
where $\iota'$ denotes the radial derivative of the rotational transform, which is directly related to the negative shear.
If we go to a second flux-tube with $\phi_2=\phi-\phi_0$ we obtain
\begin{align*}
{\bf k}_{\bot}&= k_{\psi,2}\nabla \psi+ k_{\alpha,2}\nabla \alpha_2\\
&= k_{\psi,2}\nabla \psi+ k_{\alpha,2} \left(\nabla \theta -\iota \nabla \phi -\iota' (\phi-\phi_0)\nabla \psi\right),
\end{align*}
where we find that the last term $\iota'\phi_0\nabla \psi$ can be combined with $k_{\psi,2}$ to yield $k_{\psi,1}$ again. Thus, choosing a second flux-tube corresponds to choosing a second $k_{\psi}$, even though the variation will be small if the shear $\iota'$ is small, as in the QIPC and Wendelstein 7-X. Therefore, setting $k_{\psi}=0$ will be a good assumption in W7-X and QIPC even for flux-tubes other than the one with $\alpha=0$.
Among the flux-tubes simulated in this paper, the one of NCSX with $\alpha=\pi/3$ is the only one where this choice might not lead to the highest growth rate. This is due to the high global shear of NCSX.\\
Therefore only scans in the binormal wave number, normalized by the ion sound gyroradius, $k_y\rho_s$, were performed.
For this range of wave numbers, which was usually located in the interval $k_y\rho_s=0.2 \ldots 2.0$, both the density gradient and the temperature gradient -- electron or ion temperature gradient, or both where appropriate -- were varied over the interval $a/L_x=0 \ldots 3$, where $L_x=-(\mathrm{d} \ln x/\mathrm{d}r)^{-1}$ is the gradient scale length of a quantity $x$, $r$ denominates the minor radius of the device, defined as being proportional to square root of the toroidal flux, and $a$ is the value of $r$ at the plasma boundary.
In order to display the influence of the gradients on the growth rates in the stability diagrams, the wave number with the highest growth rate $\gamma$ for each combination of gradients was identified. To discriminate between the effect of changing the gradients and the wave number, the diagrams were created using only the most unstable wave number. This procedure might of course lead to rather high wave numbers being identified as the most unstable ones, even though in quasi-linear estimates it is usually the lower wave numbers that contribute most to the transport, which roughly scales as $\propto \gamma/(k_y\rho_s)^2$. One might therefore argue that the low wave numbers are the most relevant ones, but our present focus is nevertheless on linear theory and when comparing different devices we only consider the maximum linear growth rate.\\
For simulations with two kinetic species, it is instructive to analyze the electrostatic energy transfer, which is diagnosed by a novel tool in the GENE code \cite{BanonNavarro2011}. We are particularly interested in this energy transfer close to marginal stability, since there the theory from Part I predicts stabilizing electrons in quasi-isodynamic configurations. Therefore a point in parameter space close to marginality is chosen for each energy analysis. According to Eq.~(5.26) in A.~Ba\~non Navarro's thesis \cite{BanonNavarro2012}, the electrostatic energy transfer, $\partial \mathcal{E}_{\phi}/\partial t$, defined there can be given out for each species $a$ separately and equals exactly 
$$\frac{\partial \mathcal{E}_{\phi,a}}{\partial t}= -P_a=-e_a{\rm Im}\left\lbrace J_0 \hat \phi^* \left(iv_{\|}\nabla_{\|}\hat g_a-\omega_{da}\hat g_a\right)\right\rbrace,$$
as defined in Part I. We will use the notation $\Delta E_a$ for $\partial \mathcal{E}_{\phi,a}/\partial t$.
\section{Numerical results}
\label{sec:results}
\subsection{Ion temperature gradient (ITG) modes with adiabatic electrons}
\begin{figure}
\includegraphics[width=0.45 \textwidth]{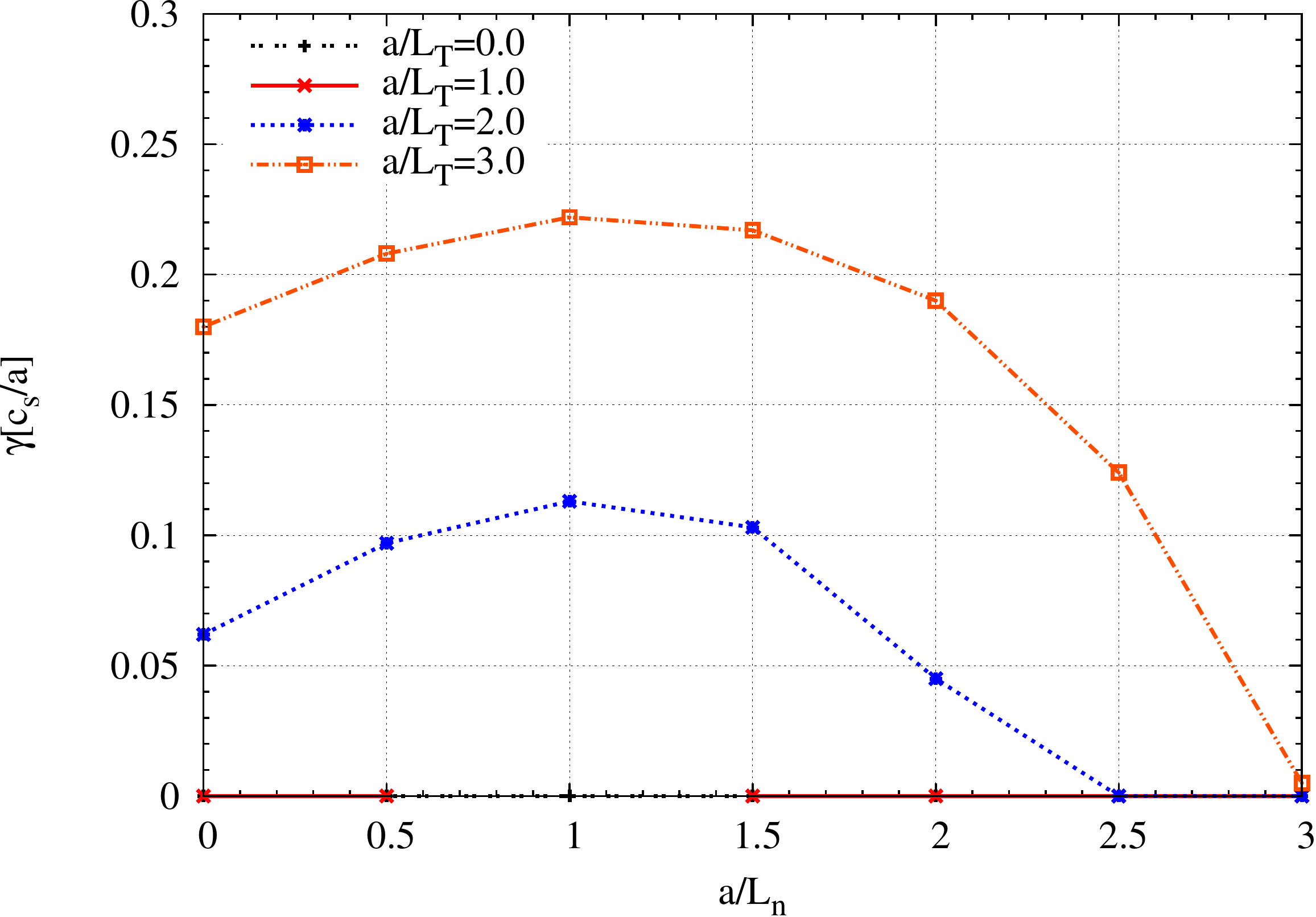}
\caption{\label{fig:ITGaTOKd3d} Growth rates for ITGs with adiabatic electrons with $k_y\rho_s=0.6$ in DIII-D.}
\end{figure}

\begin{figure}
\includegraphics[width=0.45 \textwidth]{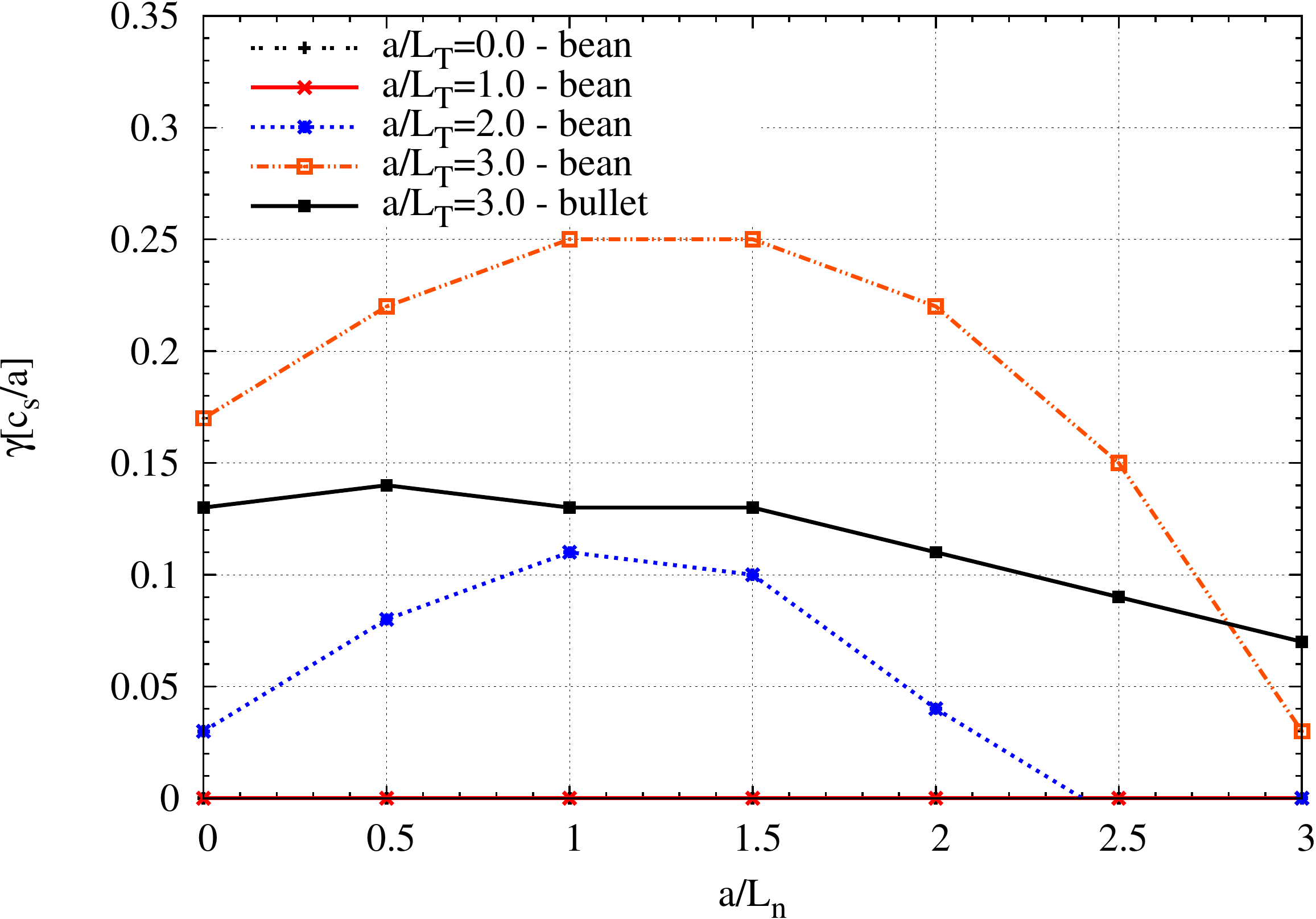}
\caption{\label{fig:ITGaNCSX} Growth rates for ITGs with adiabatic electrons with $k_y\rho_s=1.5 $ in NCSX.}
\end{figure}

\begin{figure}
\includegraphics[width=0.45 \textwidth]{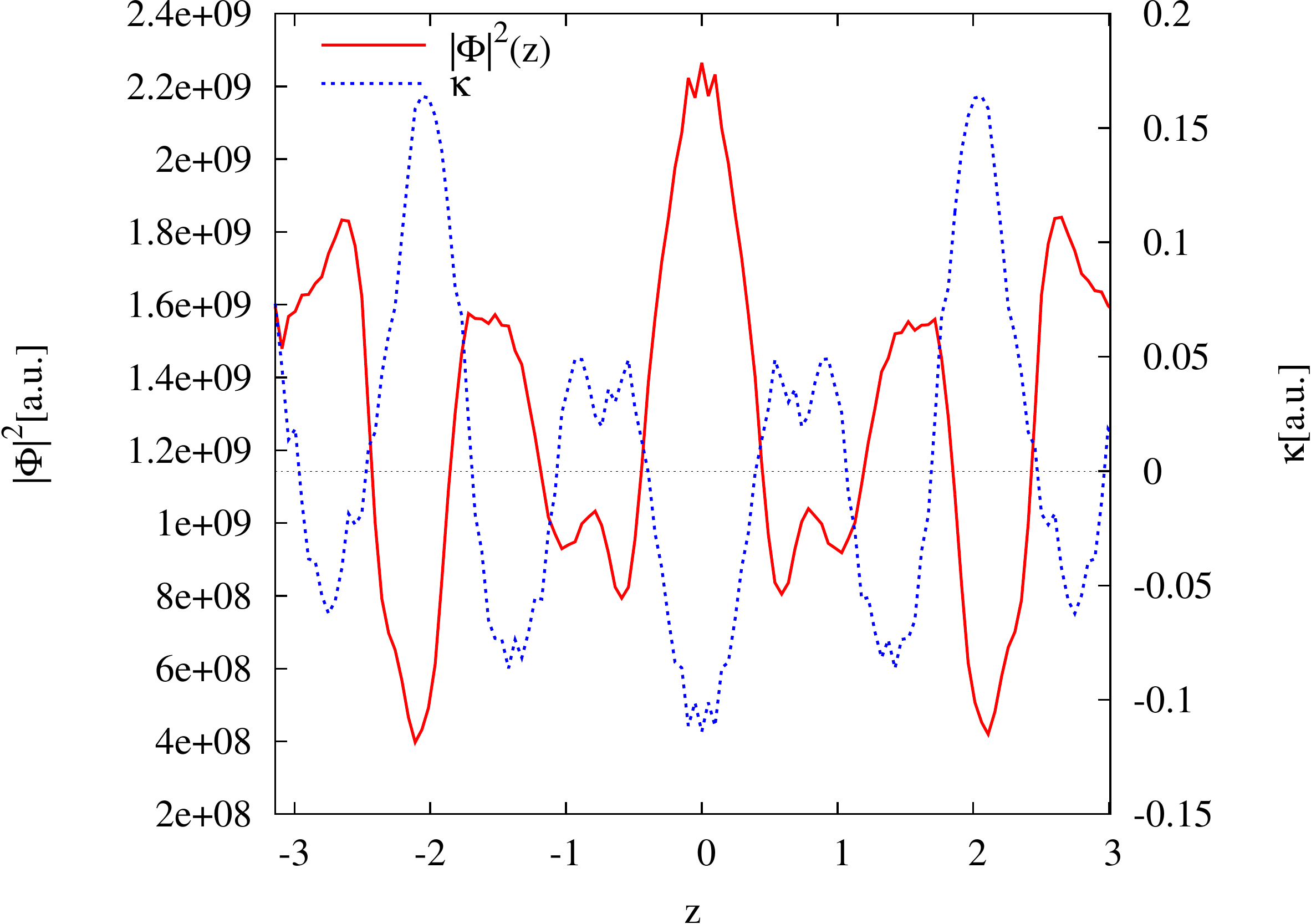}
\caption{\label{fig:ITGaW7Xphi}Typical mode structure of ITGs with adiabatic electrons in the bean flux-tube of W7X}
\end{figure}

Even though simulations using adiabatic electrons and only varying the ion temperature gradient $\nabla T_i$ and the density gradient $\nabla n$ have been performed many times in the past, even in stellarator geometry \cite{Kornilov2004a,Ferrando-Margalet2007,Watanabe2007,Nunami2011}, they were repeated here in order to illuminate the influence of the geometry on the kinetic electron response.\\
In DIII-D, most of the maximum growth rates were found at $k_y\rho_s=0.6$. In Fig.\ \ref{fig:ITGaTOKd3d} the growth rates for this wave number are displayed. It can be seen that below $a/L_{T_i}=2.0$ no unstable modes are found. For values above $a/L_{T_i}=2.0$ we find a clear destabilizing influence of the temperature gradient, whereas increasing the density gradient first leads to a destabilization of the mode but later to a stabilization. The observed modes all peak in the outboard midplane and  propagate in the ion diamagnetic direction, as is typical of ITG modes in tokamaks.\\
In NCSX we have analyzed two flux tubes. Because their general behavior is very similar and only the absolute values of the maximum growth rates differ (but not the influence of the gradients), results are only presented from one flux tube. For NCSX the more unstable flux tube is the bean flux tube, and the stability diagram in Fig.\ \ref{fig:ITGaNCSX} therefore displays the results of this flux tube. We note in passing that the reduced growth rates in the bullet flux tube is probably attributed to the selection of $k_{\psi}=0$. For comparison, the curve with the highest temperature gradient from the bullet flux tube is also included. For the simulations of the bean flux tube in NCSX, the wave numbers are shifted to slightly higher values, and the most unstable mode is found at $k_y\rho_s=1.5$. Below a temperature gradient of $a/L_{T_i}=2.0$ no unstable modes are observed, and the density gradient is destabilizing when it is weak and stabilizing when it is strong, just like in a tokamak. This feature has been observed before in NCSX \cite{Baumgaertel2012b}, and it is thought that the modes tend towards the slab limit for high density gradients \cite{Jenko2001}. The modes here also peak in the bad curvature region and propagate in the ion diamagnetic direction, again as is typical of ITG modes. The absolute values of the growth rates are marginally higher than those in DIII-D.\\
The behavior of Wendelstein 7-X with respect to ITGs with adiabatic electrons is slightly different. Again the bean flux tube is the more unstable one, even though the growth rates of the triangle flux tube are only marginally lower. The most unstable mode was found at $k_y\rho_s=1.3$. Compared with DIII-D and NCSX, we find a lower critical gradient - even for $a/L_{T_i}=1.0$ there are some unstable modes. But the influence of the gradients remains the same, with a destabilizing temperature gradient and a (de)stabilizing density gradient when this gradient is (weak) strong. This behavior has been observed before in Ref.\ \cite{Xanthopoulos2007b}. The absolute values of the growth rate lie slightly above those in DIII-D and NCSX for temperature gradients $a/L_{T_i}<3.0$ and below them for $a/L_{T_i}=3.0$, but it should be remembered that the normalization (to $c_s/a$) is different in the different devices (due to different values of $a$). Also in W7-X, the observed modes can clearly be identified as ITG modes, with peaks in the mode structure, where the curvature is negative (Fig.\ \ref{fig:ITGaW7Xphi}). This mode structure is observed mostly when the density gradient is high. If the density gradient is low, only one central peak is found.\\
\begin{figure}
\includegraphics[width=0.45 \textwidth]{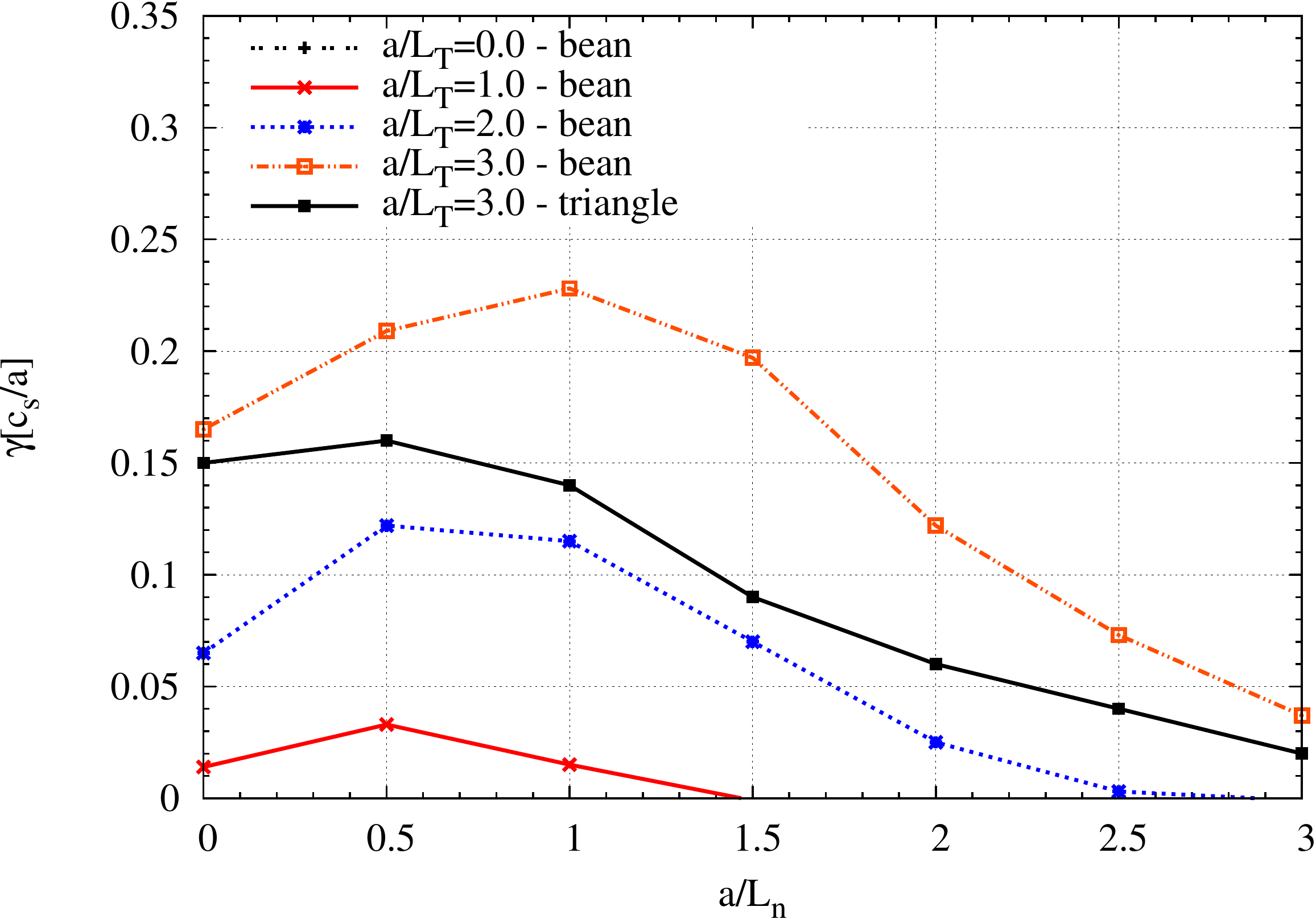}
\caption{\label{fig:ITGaW7X} Growth rates for ITGs with adiabatic electrons with $k_y\rho_s=1.3 $ in W7-X.}
\end{figure}
The QIPC stellarator has again a very similar behavior to W7-X. The bean flux tube is the most unstable one, but is only slightly more unstable than the triangle flux-tube. Also in QIPC the critical gradient lies below $a/L_{T_i}=1.0$ for the most unstable mode of $k_y\rho_s=1.8$. The modes propagate in the ion diamagnetic direction and peak where the curvature is negative. The QIPC stellarator has the lowest growth rates of all configurations examined, which might be due to the very narrow and shallow regions of bad curvature (compare for example Fig.\ \ref{fig:geometryw7xvacbean} of W7-X and Fig.\ \ref{fig:geometryqivacbean} of QIPC).
\begin{figure}
\includegraphics[width=0.45 \textwidth]{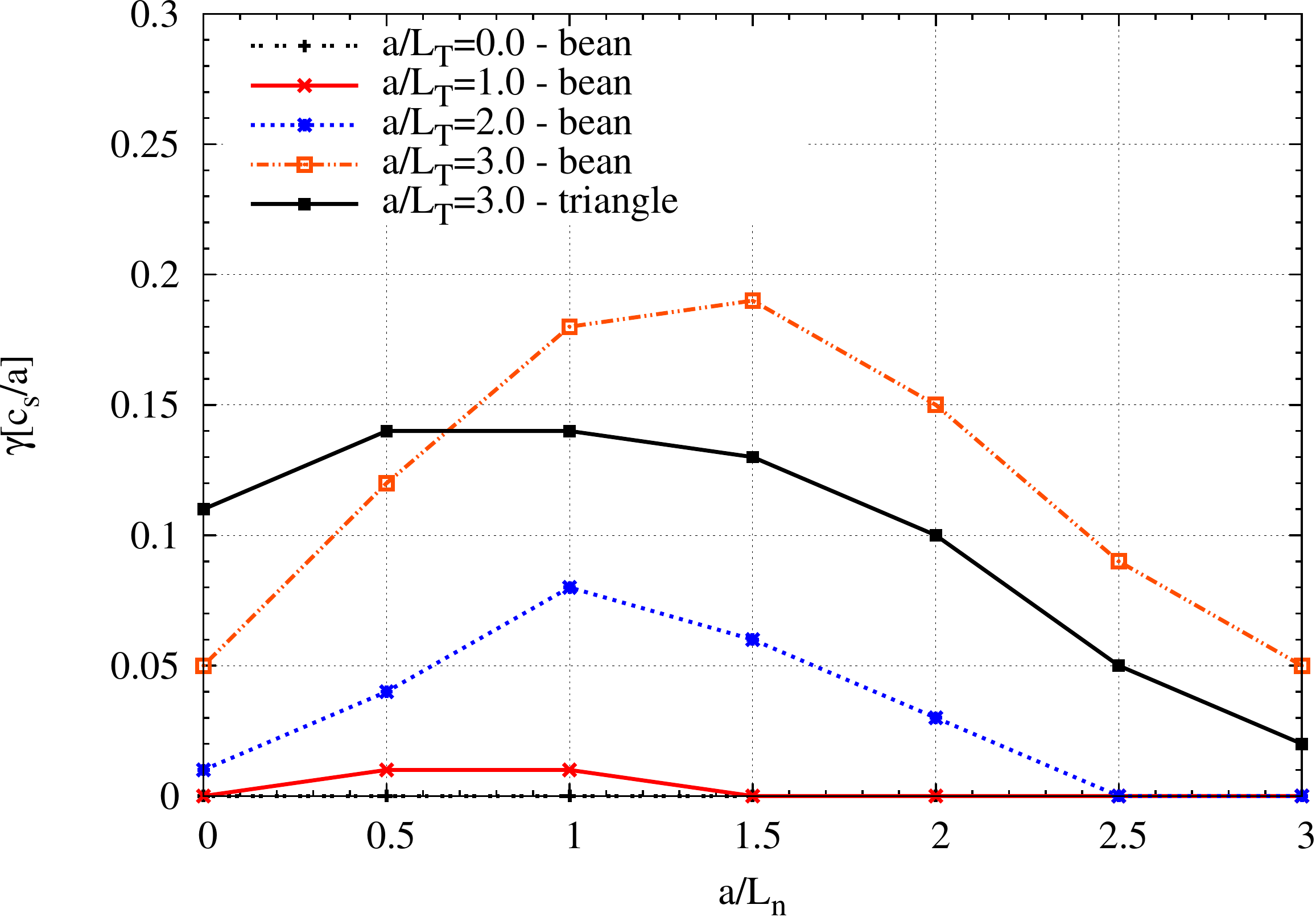}
\caption{\label{fig:ITGaQIPC} Growth rates for ITGs with adiabatic electrons with $k_y\rho_s=1.8 $ in QIPC.}
\end{figure}
\subsection{Ion temperature gradient (ITG) modes with kinetic electrons}
When the electrons are treated kinetically and their non-adiabatic response is retained, some very interesting effects emerge, especially for Wendelstein 7-X and the QIPC stellarator. At first, only the ion temperature gradient was varied, and the electron temperature gradient was set to zero. Because of quasineutrality, the ion and electron density gradients are of course equal.\\
In DIII-D a scan over $k_y$ was performed, revealing $k_y\rho_s=0.5$ as the most unstable mode. In contrast to the ITGs with adiabatic electrons, increasing the density gradient is strongly destabilizing, resulting in the maximum growth rate being a factor 3 higher than the one in the previous section. Considering the comparatively small influence of increasing the ion temperature gradient compared with the influence of the density gradient, one might suspect that the modes observed have more of a density-gradient driven TEM signature. However, even though the mode structure with its peaks in the outboard mid-plane does not reveal the nature of the mode, the propagation in the ion diamagnetic direction hints at a classical ITG mode. In addition, the energy analysis that was performed with gradients of $a/L_{T_i}=1.0$ and $a/L_{n}=0.5$ with $k_y\rho_s=0.5$, shows that the ions are the main driving species: the ratio between the energy transfer to the electrons and the ions was $\Delta E_e/\Delta E_i=-0.8$, where the minus sign indicates different energy flow directions. The ions are destabilizing and the electrons stabilizing. 
\begin{figure}
\includegraphics[width=0.45 \textwidth]{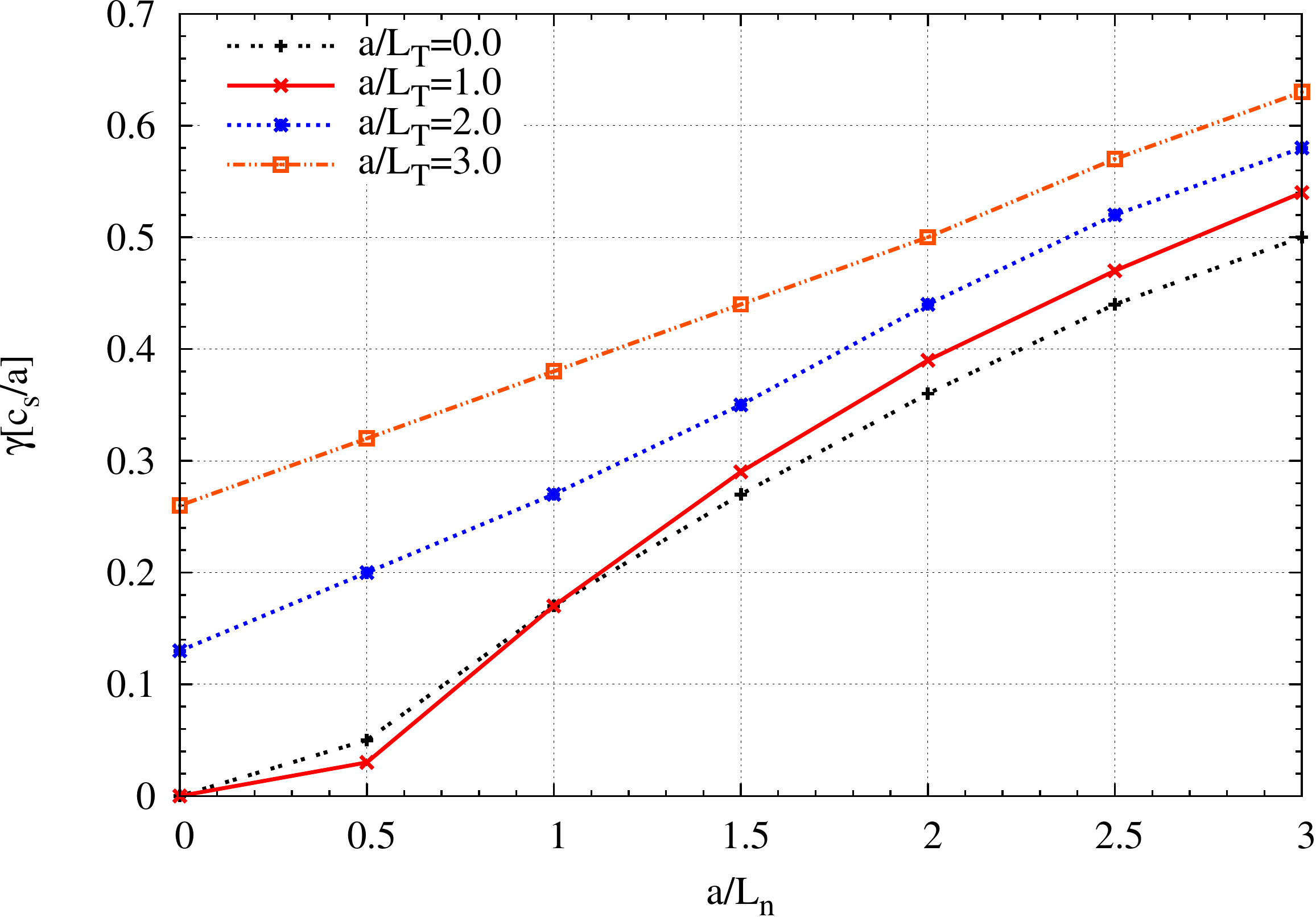}
\caption{\label{fig:ITGkTOKd3d} Growth rates for ITGs with kinetic electrons with $k_y\rho_s=0.5$ in DIII-D.}
\end{figure}

\begin{figure}
\includegraphics[width=0.45 \textwidth]{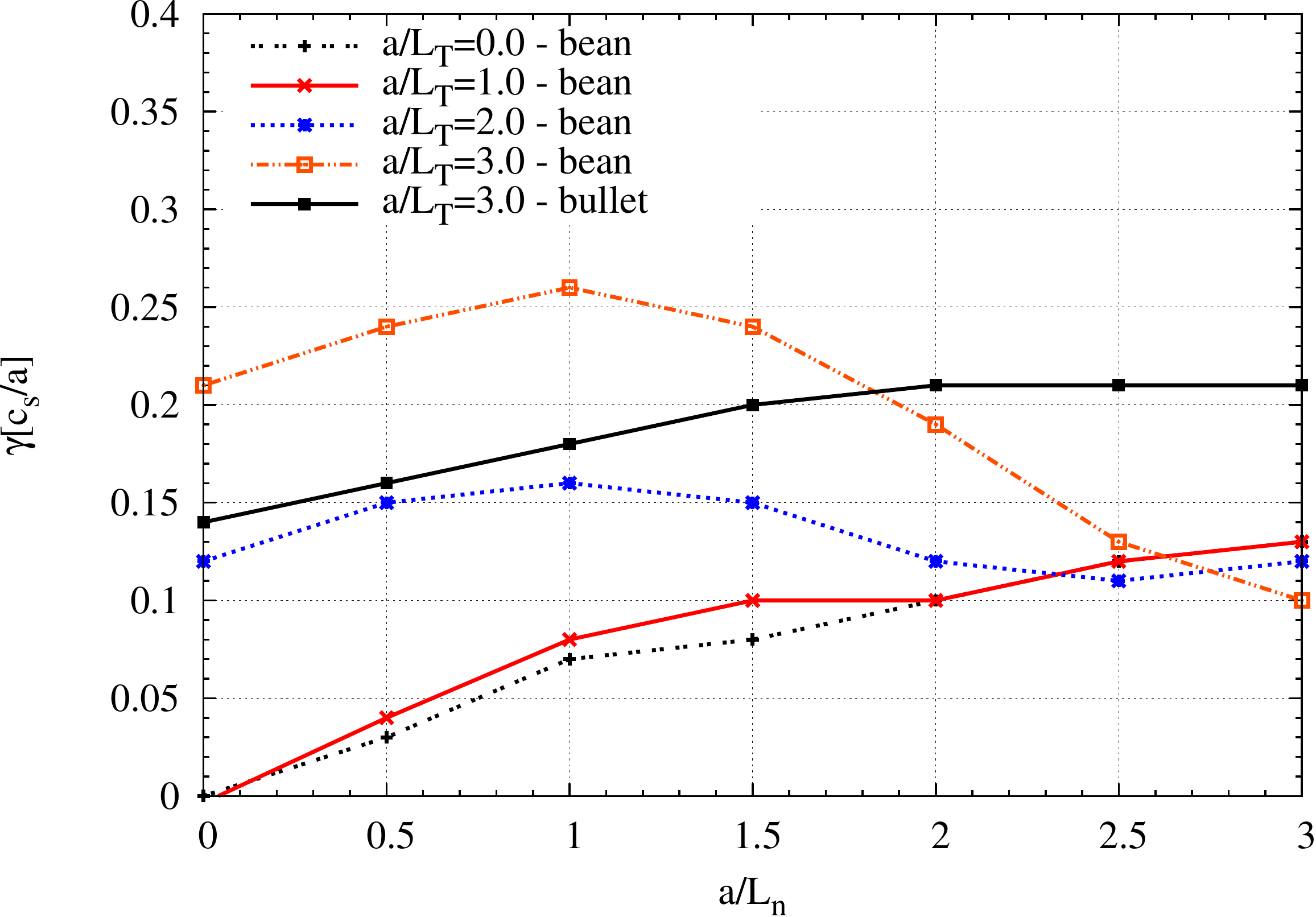}
\caption{\label{fig:ITGkNCSX} Growth rates for ITGs with kinetic electrons with $k_y\rho_s=0.7 $ in NCSX.}
\end{figure}

The influence of kinetic electrons on the stability of ITGs is similar in NCSX (Fig.\ \ref{fig:ITGkNCSX}): In the bean flux tube, which was identified as the more unstable one, the kinetic electrons have a destabilizing influence, so that there is no critical temperature gradient any more. If the ion temperature is large enough ($a/L_{T_i}=1$ and above), the modes propagate in the ion diamagnetic direction, but for small temperature gradients they propagate in the electron direction, as is typical of TEMs. The fact that the curve with $a/L_{T_i}=0$ increases monotonically with increasing density gradient (rather than decreasing above a certain point, as it is the case for classical ITGs) also suggests a density-gradient-driven TEM. This transition occurs at about $\eta_i=L_n/L_{T_i}\leq 1$. Since the focus of this section lies on ITGs, however, the energy analysis was performed for the parameters  $a/L_{T_i}=1.0$, $a/L_{n}=0.5$ and $k_y\rho_s=0.7$ (which was also identified as the most unstable wave number). Just as in DIII-D, the ions provide the main drive and the electrons stabilize the mode slightly, with the ratio of the energies again being $\Delta E_e/\Delta E_i=-0.8$. It can thus be concluded that these modes, which exist for high $\eta_i$ and are destabilized by increasing the ion temperature gradient, are indeed ITG modes. Similar results were obtained in Ref.~\cite{Baumgaertel2012b}, especially with respect to the destabilizing temperature gradient. In Ref.~\cite{Baumgaertel2013}, the relative stability of NCSX compared with a shaped tokamak is confirmed.

\begin{figure}
\includegraphics[width=0.45 \textwidth]{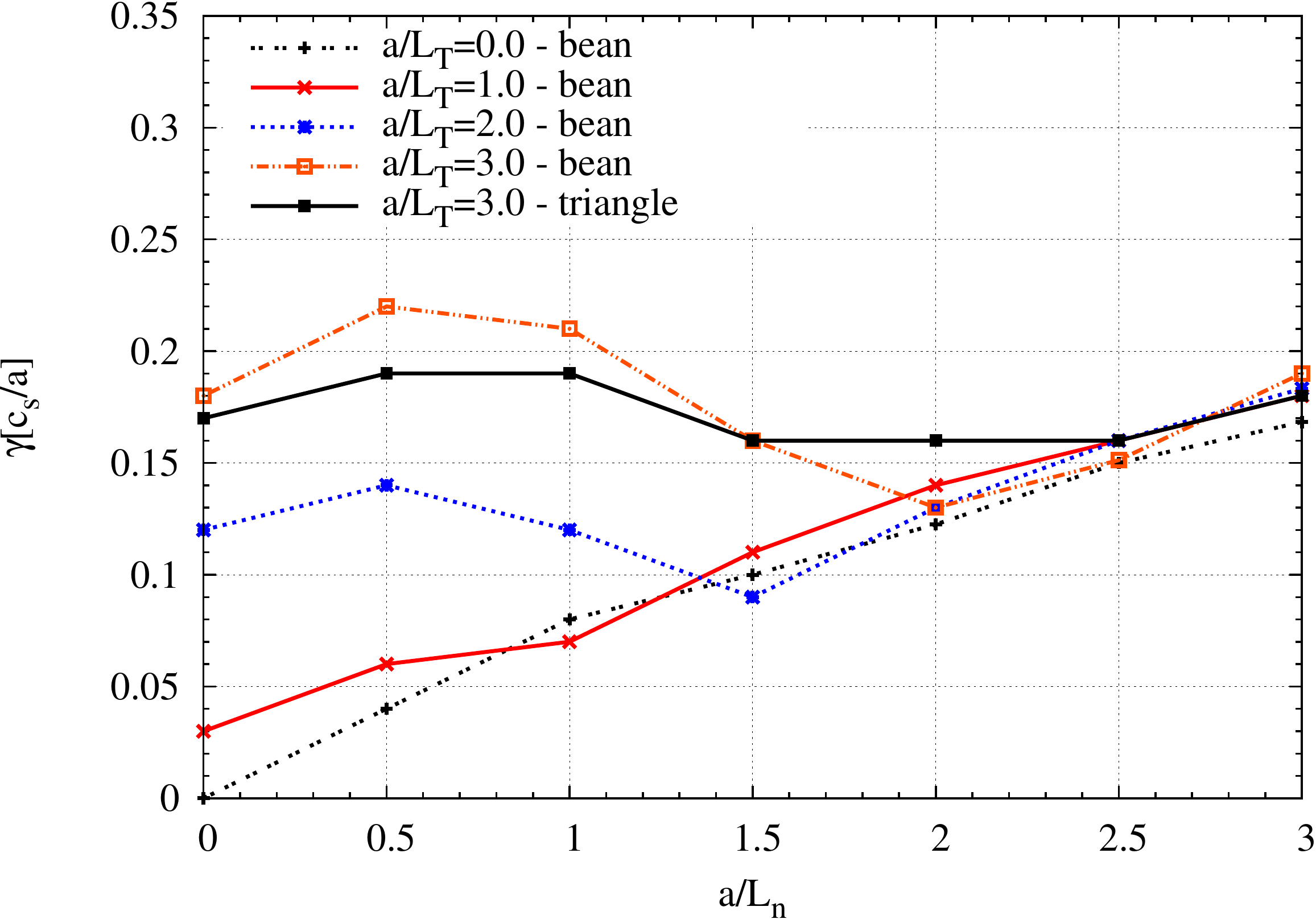}
\caption{\label{fig:ITGkW7X} Growth rates for ITGs with kinetic electrons with $k_y\rho_s=0.8 $ in W7-X.}
\end{figure}

\begin{figure}
\includegraphics[width=0.45 \textwidth]{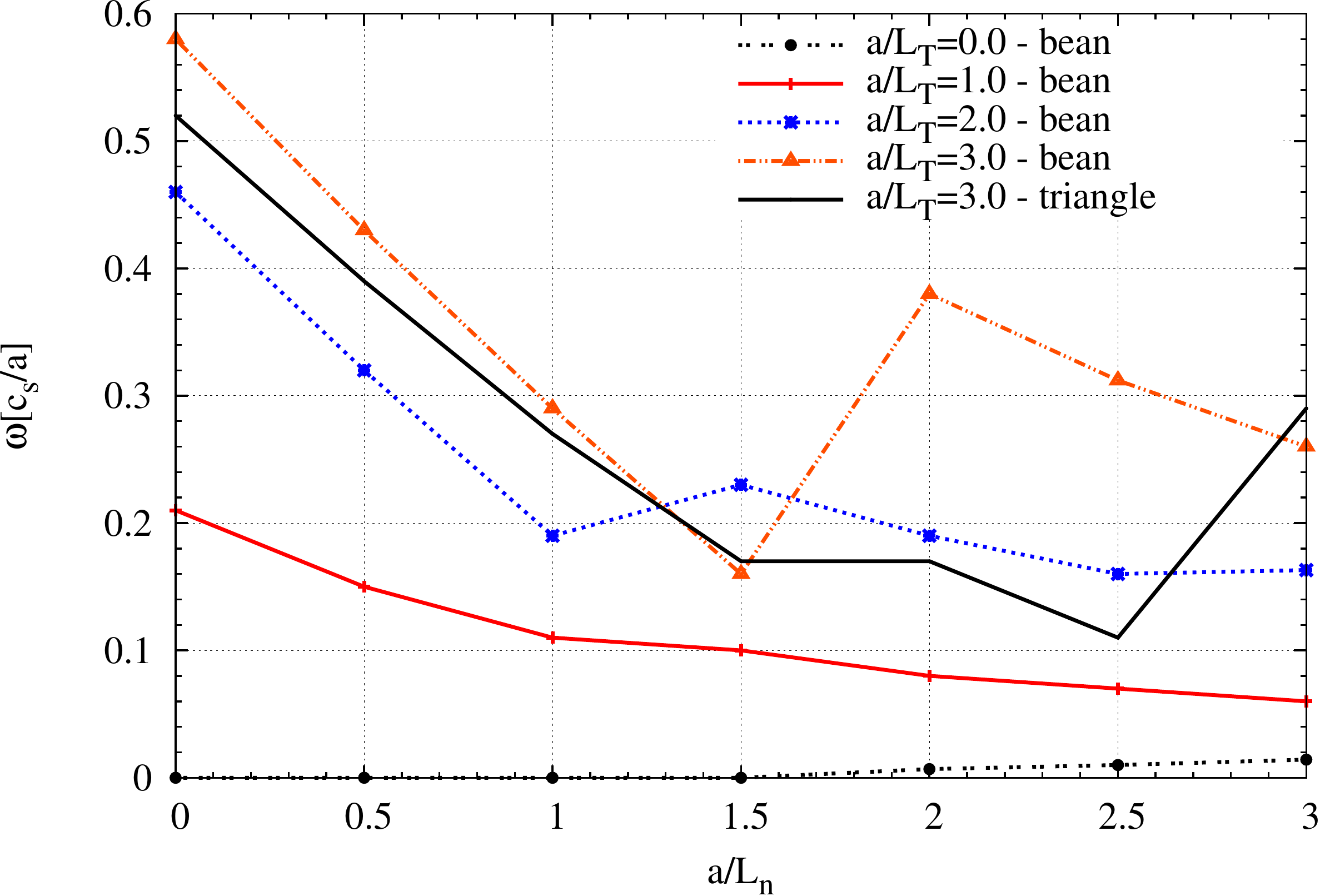}
\caption{\label{fig:ITGkW7Xomega} Real frequencies for ITGs with kinetic electrons with $k_y\rho_s=0.8 $ in W7-X.}
\end{figure}

\begin{figure}
\includegraphics[width=0.45 \textwidth]{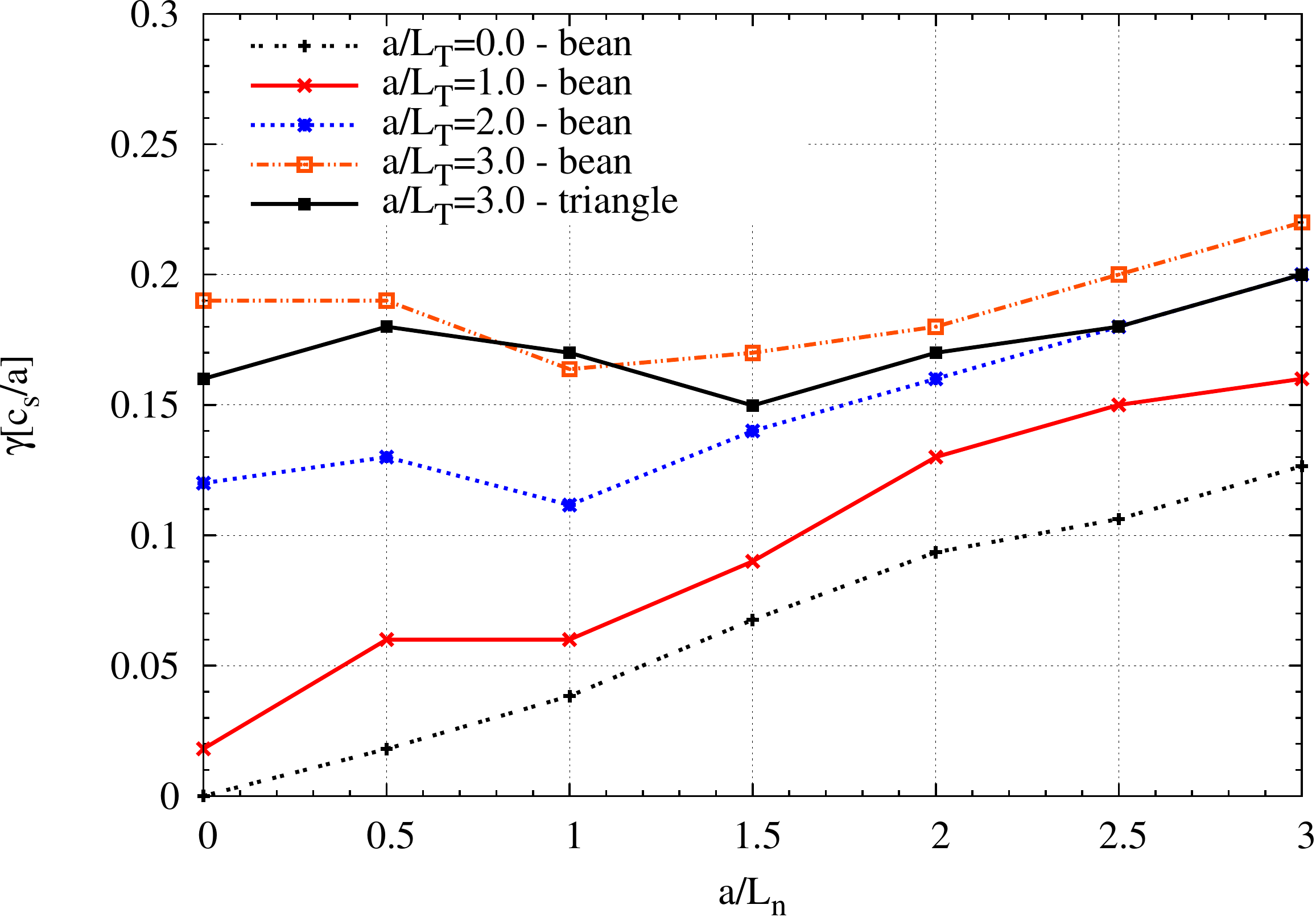}
\caption{\label{fig:ITGkQIPC} Growth rates for ITGs with kinetic electrons with $k_y\rho_s=0.6 $ in QIPC.}
\end{figure}

The transition between ITG modes and density-gradient driven-modes is even more pronounced in Wendelstein 7-X (Fig.\ \ref{fig:ITGkW7X}). Even though the $k_y$-spectra look very similar in both flux tubes, the growth rates of the bean-flux tube are slightly higher. The $k_y\rho_s=0.8$ wave number was identified as the most unstable one. For low density gradients, a destabilizing effect of the ion temperature gradient can be observed. In this region, the modes propagate in the ion diamagnetic direction. Also the mode structures, with the maxima coinciding with the bad curvature regions, indicate classical ITGs. If the density gradient is high enough, however, a mode transition occurs (Fig.\ \ref{fig:ITGkW7Xomega}): not only does the influence of the temperature gradient vanish, but also the mode structure shifts towards the magnetic wells. These modes can thus be classified as trapped-particle modes, even though the direction of propagation is still in the ion diamagnetic direction, unlike the classical TEM. For the energy analysis, the first ITG mode appears at $a/L_{T_i}=2.0$ and $a/L_{n}=0.5$ and $k_y\rho_s=0.8$, and it was found that the electrons stabilize this mode, with the ratio of energy transfer being $\Delta E_e/\Delta E_i=-0.8$.\\ 
In QIPC, the density-gradient-driven modes become even more dominant. For the bean flux tube -- again identified as the more unstable one -- the transition between ITG and trapped-particle mode occurs at lower density gradients than for W7-X  (Fig.\ \ref{fig:ITGkQIPC}). Both types of mode propagate in the ion diamagnetic direction, but the mode structure reveals clear differences between the ITG and the trapped-particle mode, just as in W7-X.  This might be due to the generally lower growth rates of ITGs (see the previous subsection), so that the trapped-particle mode, which is similarly unstable in both configurations, becomes dominant at lower density gradients. A justified question would be whether the trapped-particle modes that are observed both in W7-X as well as in QIPC could be classified as trapped-ion modes. In our understanding, however, an ordinary trapped-ion mode requires the trapped character of the ion orbits to be important, which is not the case here since the frequency of the mode exceeds the ion bounce frequency by far. Moreover, kinetic electrons are necessary for this mode to exist, which should rule out the classification as a trapped-ion mode. For QIPC, both mode types are analyzed with respect to the energy transfer. For the ITGs, the parameters chosen were $a/L_{T_i}=2.0$, $a/L_{n}=0.0$ and $k_y\rho_s=0.6$. Again, the electrons were found to be drawing energy from the mode, with the ratio of energy flux being $\Delta E_e/\Delta E_i=-0.7$. The same result was obtained for the trapped-particle mode at gradients of $a/L_{T_i}=2.0$ and $a/L_{n}=2.0$. The most unstable wave number was found at $k_y\rho_s=0.6 $.
\subsection{Trapped-electron modes (TEM)}
\begin{figure}
\includegraphics[width=0.45 \textwidth]{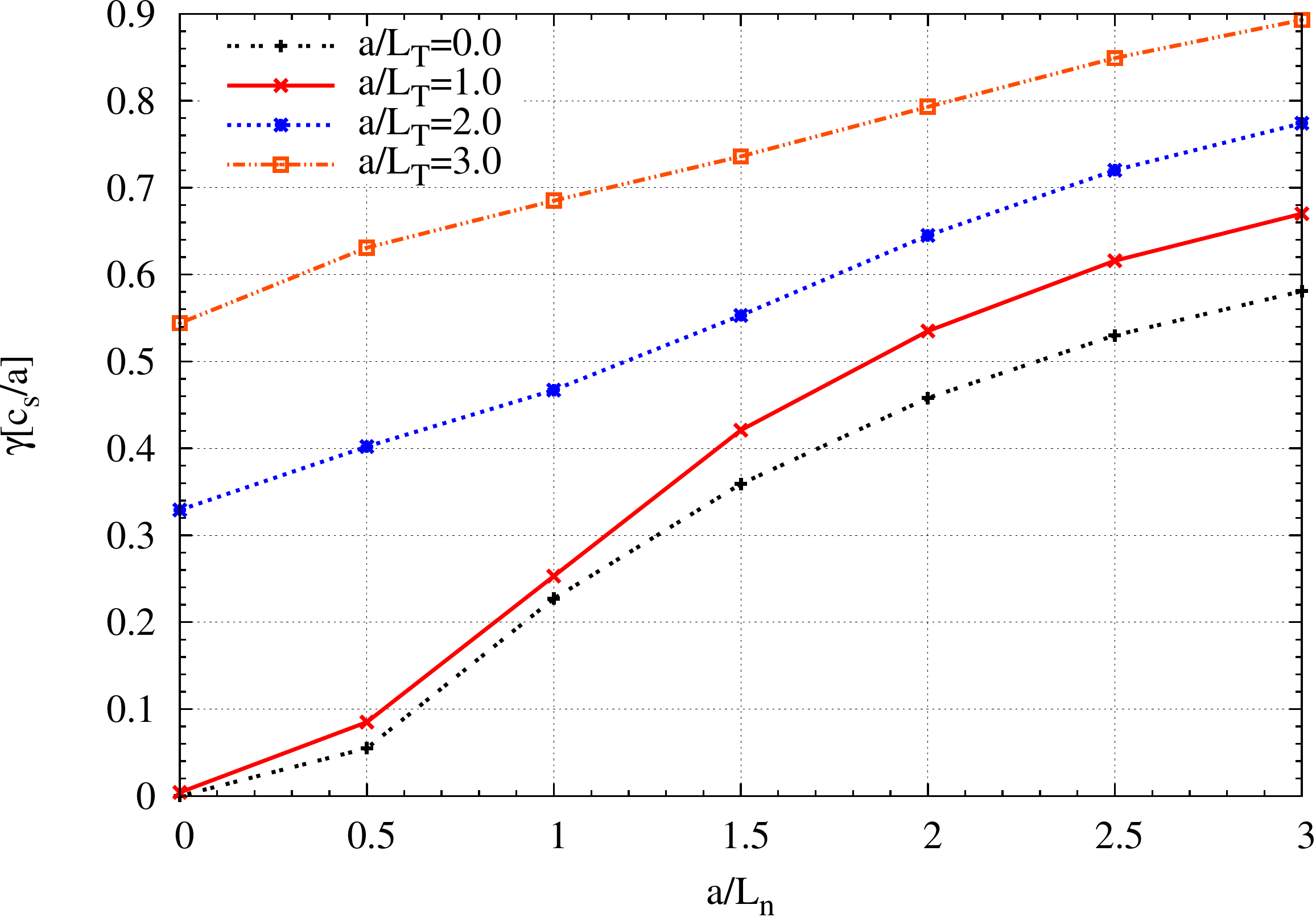}
\caption{\label{fig:TEMTOKd3d} Growth rates for TEMs with $k_y\rho_s=1.5$ in DIII-D.}
\end{figure}

\begin{figure}
\includegraphics[width=0.45 \textwidth]{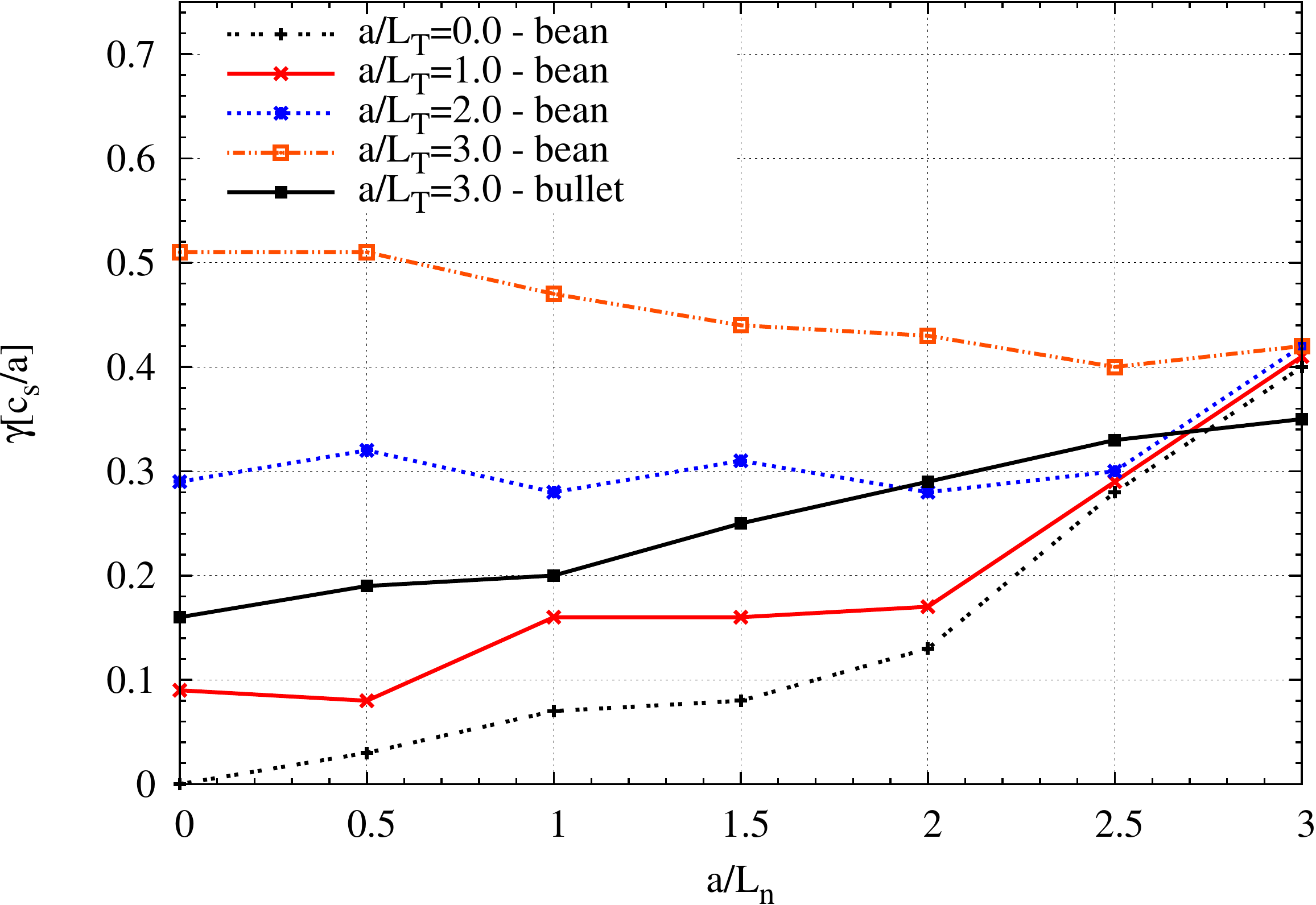}
\caption{\label{fig:TEMNCSX} Growth rates for TEMs with $k_y\rho_s=2.5 $ in NCSX.}
\end{figure}

\begin{figure}
\includegraphics[width=0.45 \textwidth]{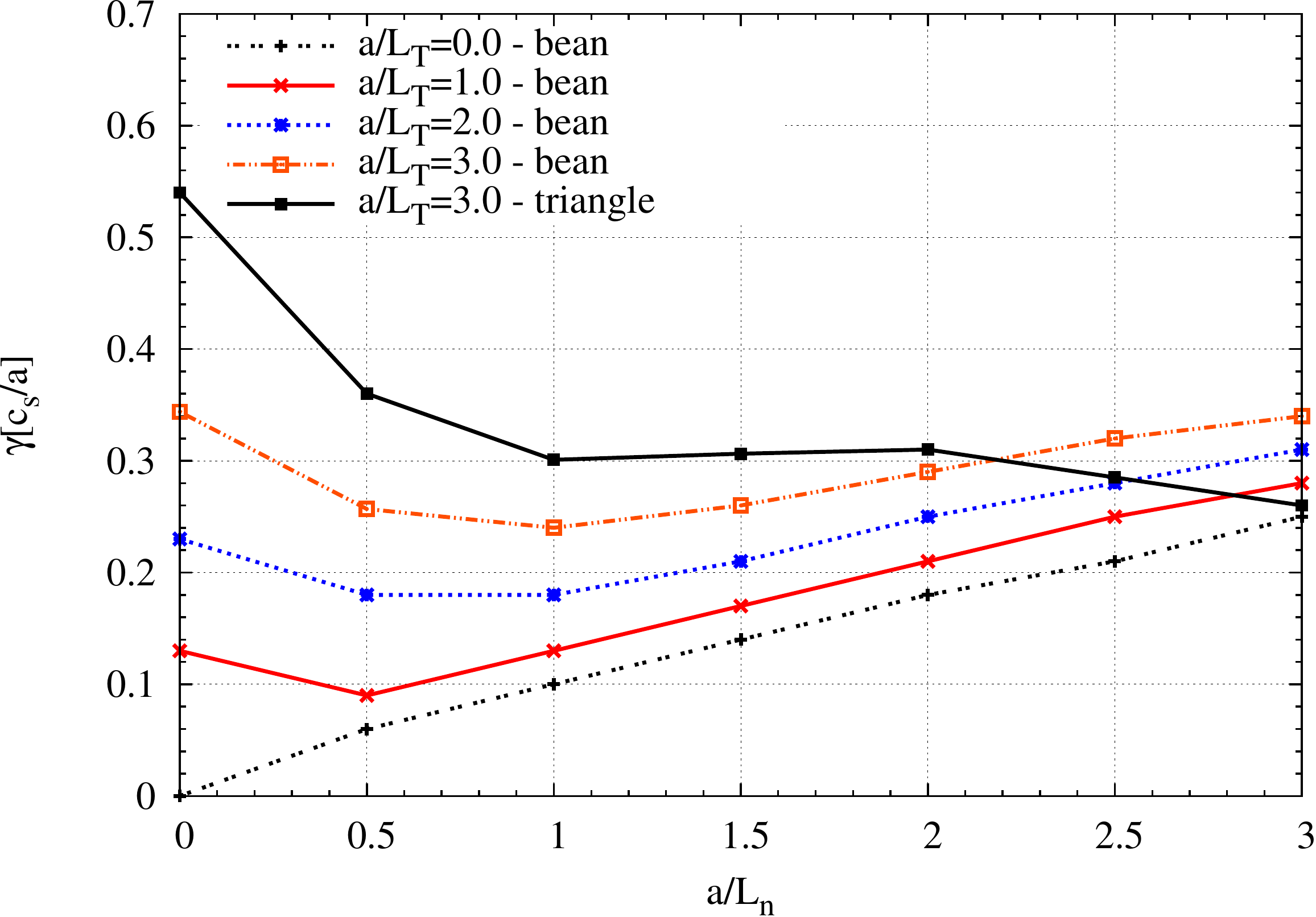}
\caption{\label{fig:TEMW7X} Growth rates for TEMs with $k_y\rho_s=1.6 $ in W7-X.}
\end{figure}

\begin{figure}
\includegraphics[width=0.45 \textwidth]{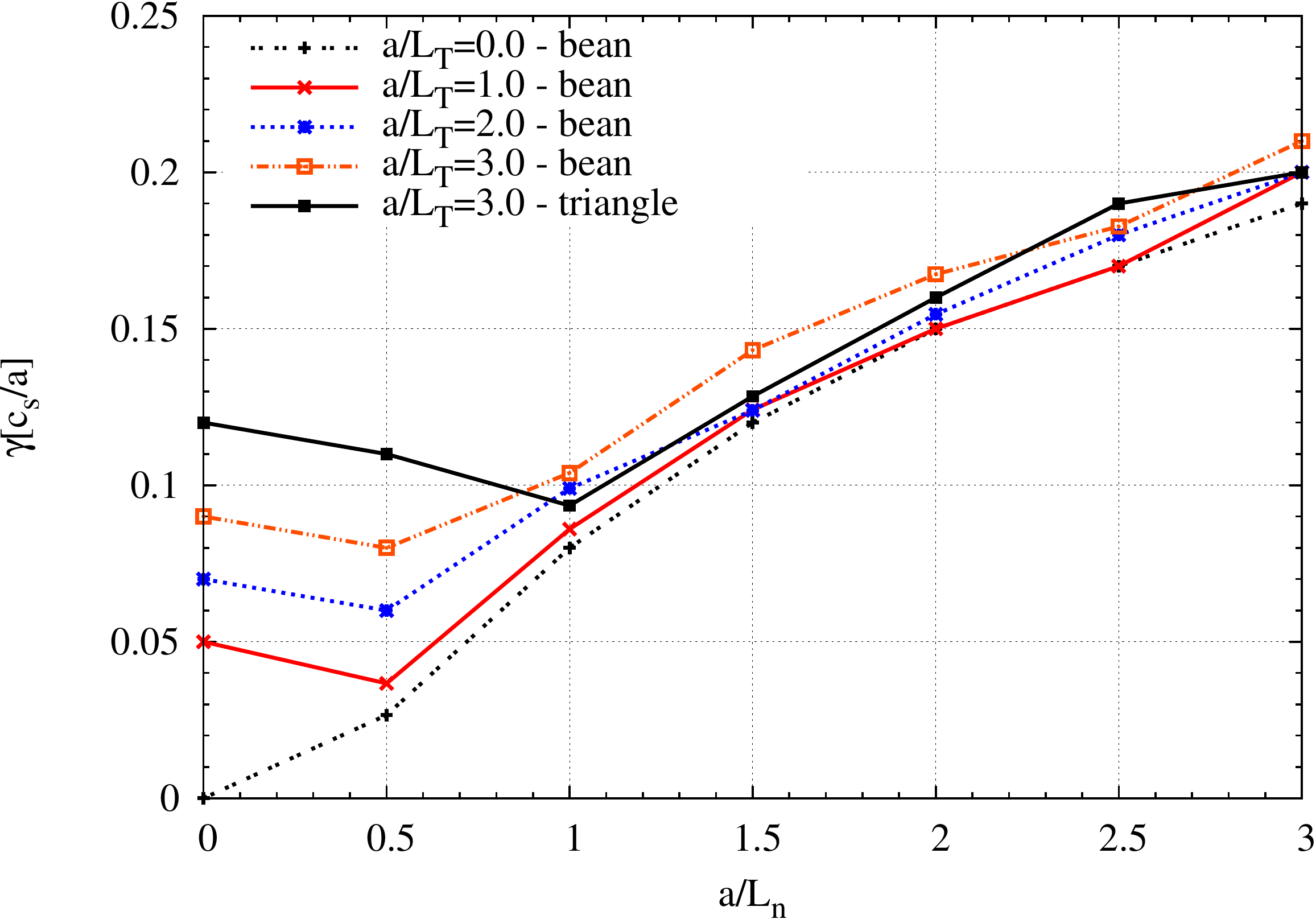}
\caption{\label{fig:TEMQIPC} Growth rates for TEMs with $k_y\rho_s=0.9 $ in QIPC.}
\end{figure}

\begin{figure}
\includegraphics[width=0.45 \textwidth]{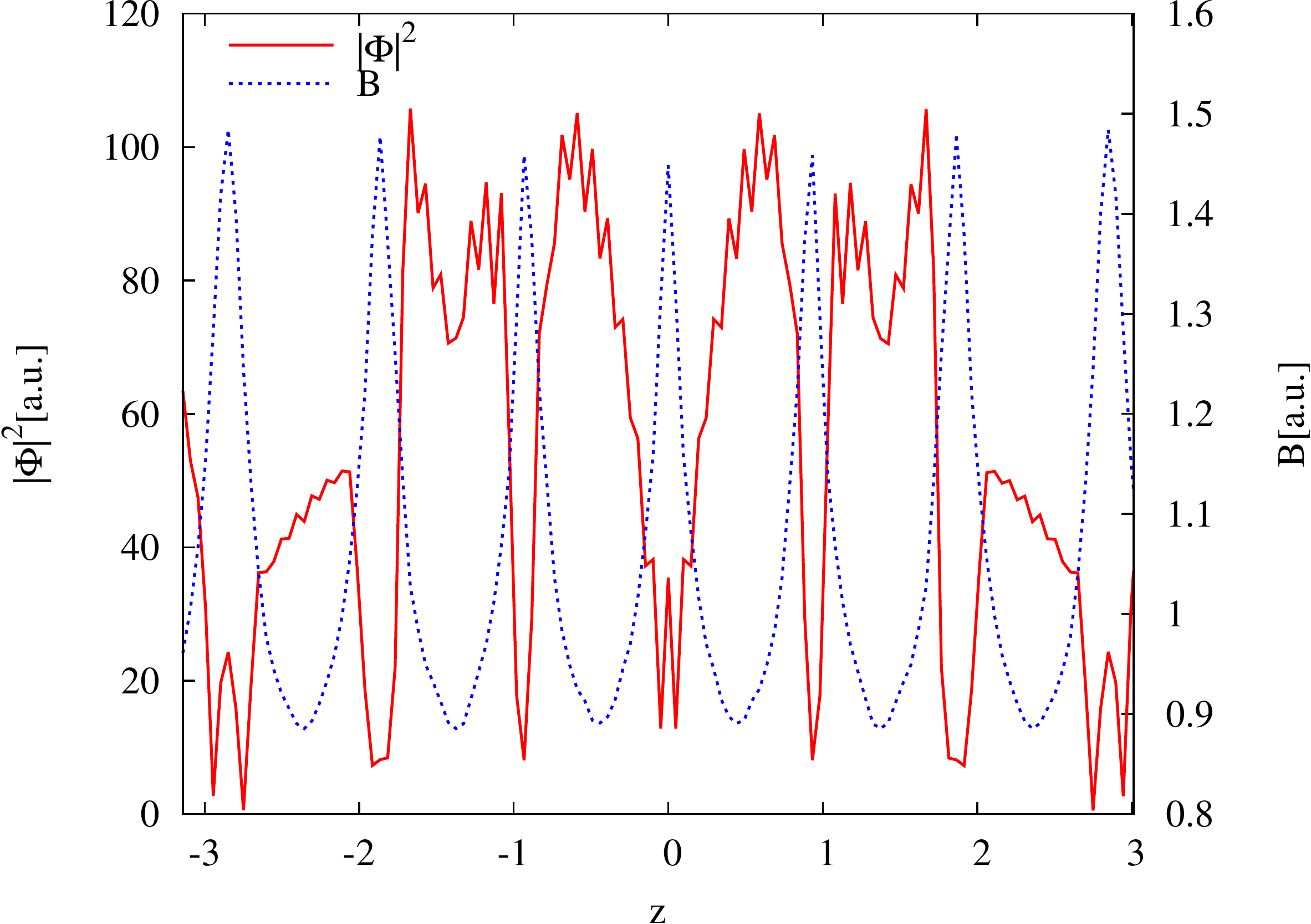}
\caption{\label{fig:TEMQIPCphi}Typical mode structure of a trapped-particle mode in the bean flux-tube of QIPC}
\end{figure}

In order to isolate trapped-electron modes, the ion temperature gradient was set to zero, so that only the electron temperature and density gradients were varied.\\
In DIII-D the most unstable wave number for the trapped-electron modes was found to be $k_y\rho_s=1.5$. Increasing either the density gradient or the electron temperature gradient is always found to be destabilizing (Fig.\ \ref{fig:TEMTOKd3d}). For low density gradients, one finds a critical temperature gradient between $a/L_{Ti}=1$ and $a/L_{Ti}=2$. At higher density gradients, however, the destabilization through the density gradient becomes so large that an unstable mode can be found even at vanishing temperature gradient. The mode frequency changes monotonically with increasing gradient, so that the modes propagate in the electron diamagnetic direction when the gradients are small, and in the opposite direction for high density gradients. This behaviour is nothing unusual for TEMs in tokamaks and has been observed before (e.g. in Ref.\ \cite{Kammerer2008}), even though propagation in the electron diamagnetic direction is more common \cite{Gorler2008, Pueschel2008, Pueschel2010, Bravenec2011}. The energy analysis at gradients of $a/L_{T_e}=1.0$ and $a/L_{n}=0.5$ (this time a slightly different wave number of $k_y\rho_s=1.1$  was selected) reveals that the modes observed here are indeed electron-driven and thus ordinary TEMs. The energy transfer from the electrons to the mode is also very large compared with the energy drawn from the mode by the ions, with a ratio of $\Delta E_e/\Delta E_i=-10$.\\
NCSX is found to be more stable than DIII-D (Fig.\ \ref{fig:TEMNCSX}). The bean flux tube turned out to be the significantly more unstable one, with the growth rate peaking at $k_y\rho_s=2.5 $. It is remarkable how the modes change with increasing density gradient, from a strongly temperature-gradient-driven mode for low density gradients to a purely density-gradient-driven mode, where the temperature gradient has almost no destabilizing influence.
All modes observed propagate in the electron diamagnetic direction, and the energy analysis at $a/L_{T_e}=1.0$ and $a/L_{n}=0.5$ reveals a strong destabilizing effect by the electrons with a ratio of $\Delta E_e/\Delta E_i=-12$. The modes observed can thus be identified as classical electron-driven TEMs.\\
Wendelstein 7-X exhibits somewhat different behaviour. The most unstable wave number was found at $k_y\rho_s=1.6 $. As can be seen in Fig.\ \ref{fig:TEMW7X}, the instability sets in already at very low gradients, but the growth rates remain lower than for DIII-D or NCSX over most of parameter space. 
\begin{figure}
\includegraphics[width=0.45 \textwidth]{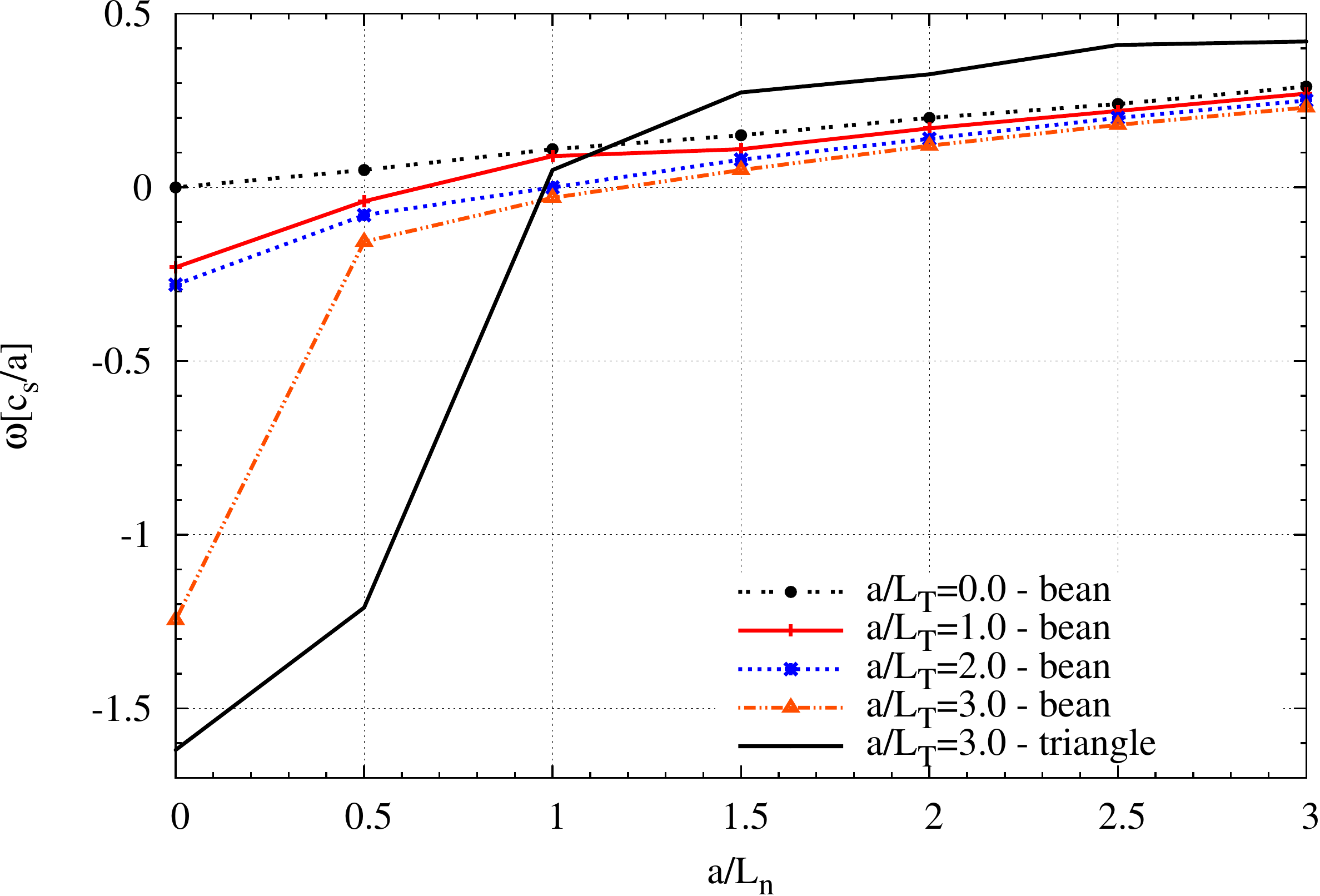}
\caption{\label{fig:TEMW7Xomega} Real frequencies for TEMs with $k_y\rho_s=1.6 $ in W7-X.}
\end{figure}
Most of the modes observed -- especially in the region of strong density gradients, where the curves for different temperature gradients are parallel to each other -- are trapped-particle modes, i.e. their mode amplitude peaks in the magnetic wells. But these modes propagate in the ion diamagnetic direction (see Fig.\ \ref{fig:TEMW7Xomega}), and the energy analysis for one of these modes at gradients of $a/L_{T_e}=1.0$ and $a/L_{n}=1.0$ and $k_y\rho_s=1.6 $ shows that it is the ions, and not the electrons, that provide most of the drive, with the ratio of $\Delta E_e/\Delta E_i=0.1$. Thus, even though the gradients classically trigger electron-driven TEMs, the observed mode does not belong to this category. Since the main drive stems from the ions, a trapped-ion mode would be another option, but the mode frequencies observed exceed the ion bounce frequency considerably. Therefore, this type of instability evades standard classification. In addition to this trapped-particle instability, another mode is observed at low density gradients. This mode has a very unusual mode structure, with its main peaks in the bad-curvature region and a rather extended background structure that has its maximum at the outboard midplane. Since the ion temperature gradient is set to zero, the appearance of an ITG mode is not possible; also the mode propagates in the electron diamagnetic direction. 
\begin{figure}
\begin{center}
\includegraphics[width=0.45 \textwidth]{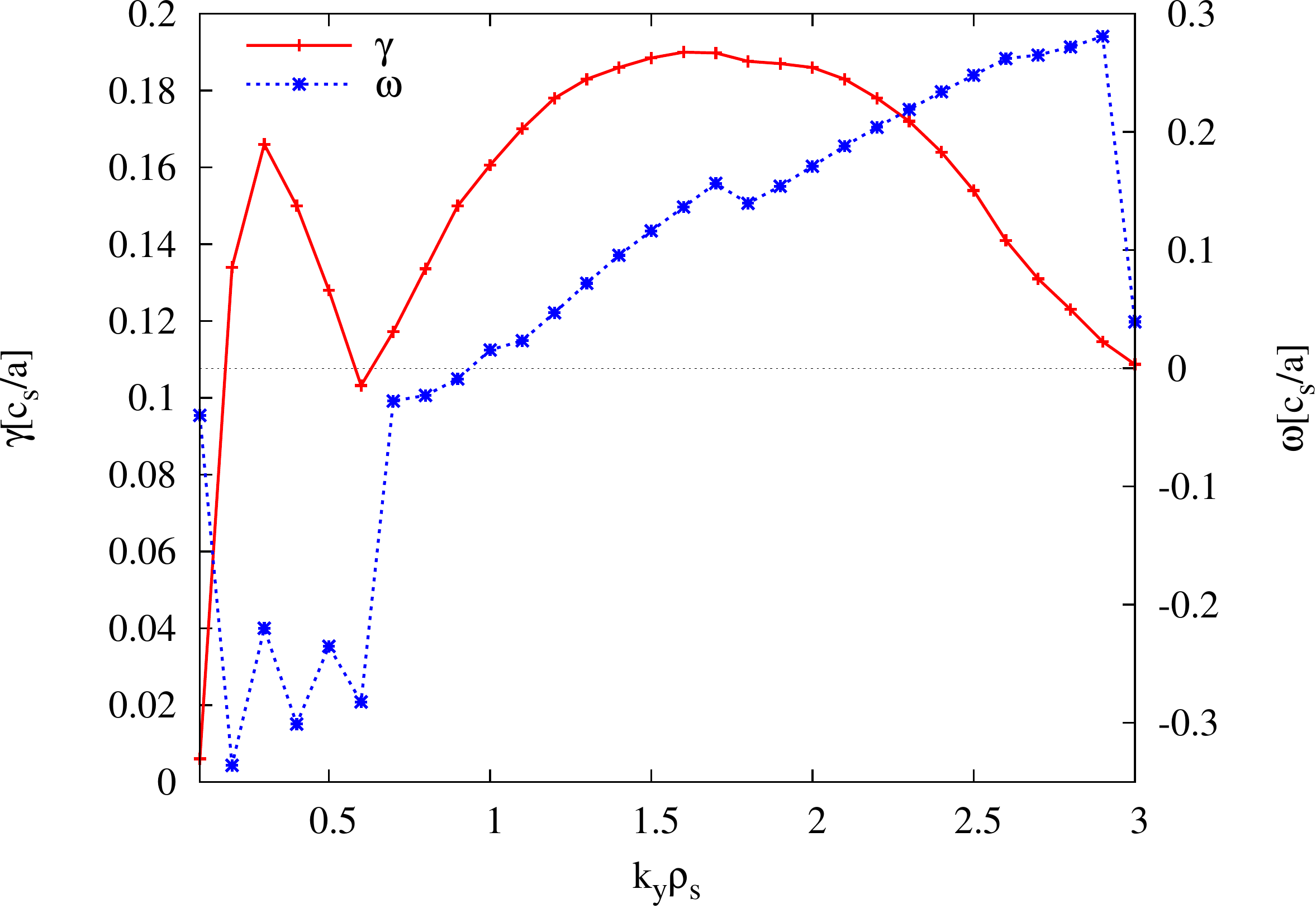}
\end{center}
\caption{Scan of $k_y\rho_s$ in the triangle flux-tube of W7-X at $a/L_n=2.0$ and $a/L_{T_e}=1.0$.}
\label{fig:batmodekyscan}
\end{figure}
For these modes to be electron temperature gradient modes (ETG) the wave numbers are not large enough compared with the high wave numbers observed for ETGs in tokamaks \cite{Jenko2000}. This can be seen in Fig.\ \ref{fig:batmodekyscan}, where a scan over $k_y\rho_s$ was performed in the triangle flux-tube. The non-trapped-particle mode is only found up to $k_y\rho_s=0.6$, which can be seen from the negative mode frequency and also by looking at the mode structure. For higher wave numbers, this mode is stabilized and the trapped-particle mode is observed.
We find these two novel types of modes in QIPC as well (Fig.\ \ref{fig:TEMQIPC}). The fastest growing mode with $k_y\rho_s=0.9 $ exhibits a very clear destabilizing influence of the density gradient. This mode is quite similar to the one observed in W7-X, with the mode structure peaking in the magnetic wells (Fig.\ \ref{fig:TEMQIPCphi}), but its propagation is in the ion diamagnetic direction. In QIPC the mode is not only mainly driven by the ions but is actually stabilized by the electrons. The energy analysis performed at $a/L_{T_e}=1.0$ and $a/L_{n}=1.0$ and $k_y\rho_s=0.9$ yielded $\Delta E_e/\Delta E_i=-0.5$. Accordingly, this mode cannot be identified as an electron-driven TEM. Compared with W7-X, this mode is even less influenced by the temperature gradient, and its growth rates are significantly lower than those in W7-X, not to mention NCSX and DIII-D. In addition, the same non-trapped-particle mode as in W 7-X is found at low density gradients, again propagating in the electron diamagnetic direction.
\subsection{Mixed ITG-TE modes}
\begin{figure}
\includegraphics[width=0.45 \textwidth]{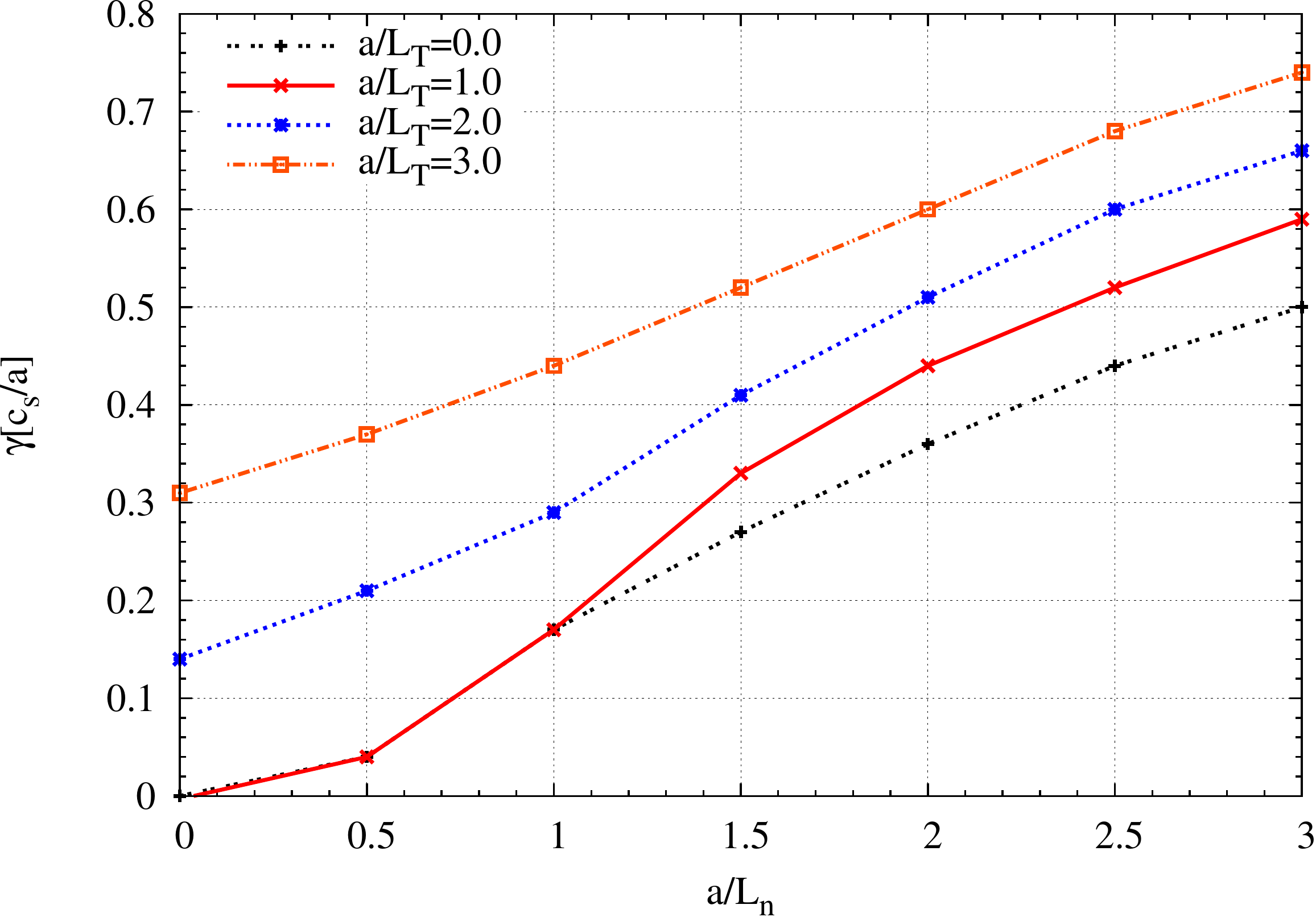}
\caption{\label{fig:ITGTEMTOKd3d} Growth rates for mixed ITG-TEMs with $k_y\rho_s=0.6$ in DIII-D.}
\end{figure}

\begin{figure}
\includegraphics[width=0.45 \textwidth]{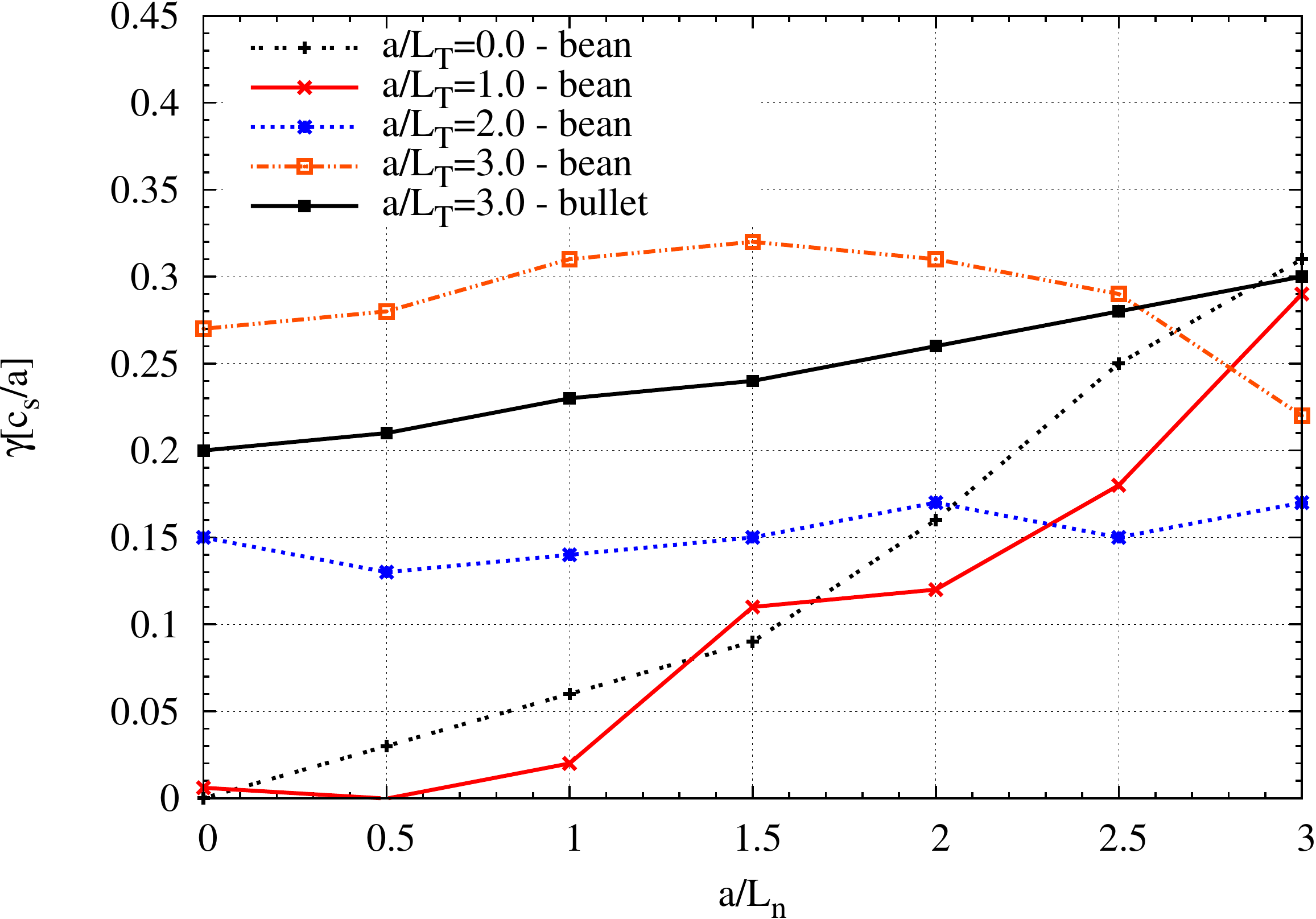}
\caption{\label{fig:ITGTEMNCSX} Growth rates for mixed ITG-TEMs with $k_y\rho_s=1.6 $ in NCSX.}
\end{figure}

In this section, we investigate instabilities where both temperature gradients are non-zero and equal, i.e. $a/L_{T_e}=a/L_{T_i}\equiv a/L_{T}$.\\ 
In DIII-D, there is only little change compared with the pure TEM case (Fig.\ \ref{fig:ITGTEMTOKd3d}), as both temperature and density gradients are destabilizing. However, the resulting growth rates are slightly reduced compared with the pure TEM case. Also, the frequency of the modes is monotonically falling, with positive frequencies $\omega>0$ for low density gradients or high temperature gradients and negative frequencies $\omega <0$ for high density gradients or low temperature gradients, in contrast to the monotonically growing frequency of the TEMs. The different signs correspond to, as we shall see, an ITG mode and a TEM. Singling out parameters of $a/L_{T}=0$ and $a/L_{n}=0.5$ and $k_y\rho_s=0.5$, where the mode frequency is in the electron diamagnetic direction, i.e. $\omega<0$, it is found that the electrons are strongly destabilizing and the ions stabilizing, with a ratio of $\Delta E_e/\Delta E_i=-10$, thus indicating an ordinary electron-driven TEM. If the gradients are chosen differently, $a/L_{T}=1.0$ and $a/L_{n}=0.5$ and $k_y\rho_s=0.5$, where the mode propagates in the ion diamagnetic direction, we find, just as expected for an ITG mode, that the ions is the destabilizing species and the electrons are stabilizing, with a ratio of $\Delta E_e/\Delta E_i=-0.7$. The maximum growth rate was found at $k_y\rho_s=0.6$.\\
Also in NCSX this general behaviour prevails (Fig.\ \ref{fig:ITGTEMNCSX}), but the growth rates are slightly reduced compared with the TEM case. The highest growth rates were found for $k_y\rho_s=1.6 $. In the more unstable bean flux tube, the modes observed were both of the ITG and TEM type, with the mode frequency alternating between the ion and electron diamagnetic directions. A tendency towards the ion diamagnetic frequency could be observed when increasing the temperature gradients, and it could thus be concluded that ITGs are dominant for high temperature gradients. This is supported by the observation that, for high temperature gradients, the density gradient has a slightly stabilizing influence, just as observed in the case of kinetic electron ITGs in NCSX (Fig.\ \ref{fig:ITGkNCSX}). For low temperature gradients, however, for example at the point where the energy analysis was performed with gradients of $a/L_{T}=1.0$ and $a/L_{n}=0.5$ and $k_y\rho_s=1.6$, the electrons were found to be the destabilizing species. The energy transfer ratio was $\Delta E_e/\Delta E_i=-1.0$. Also in these cases, increasing the density gradient continuously destabilizes the modes. A very similar behaviour was already observed for NCSX in Fig.\ 20 of Ref.~\cite{Baumgaertel2012b}.

\begin{figure}
\includegraphics[width=0.45 \textwidth]{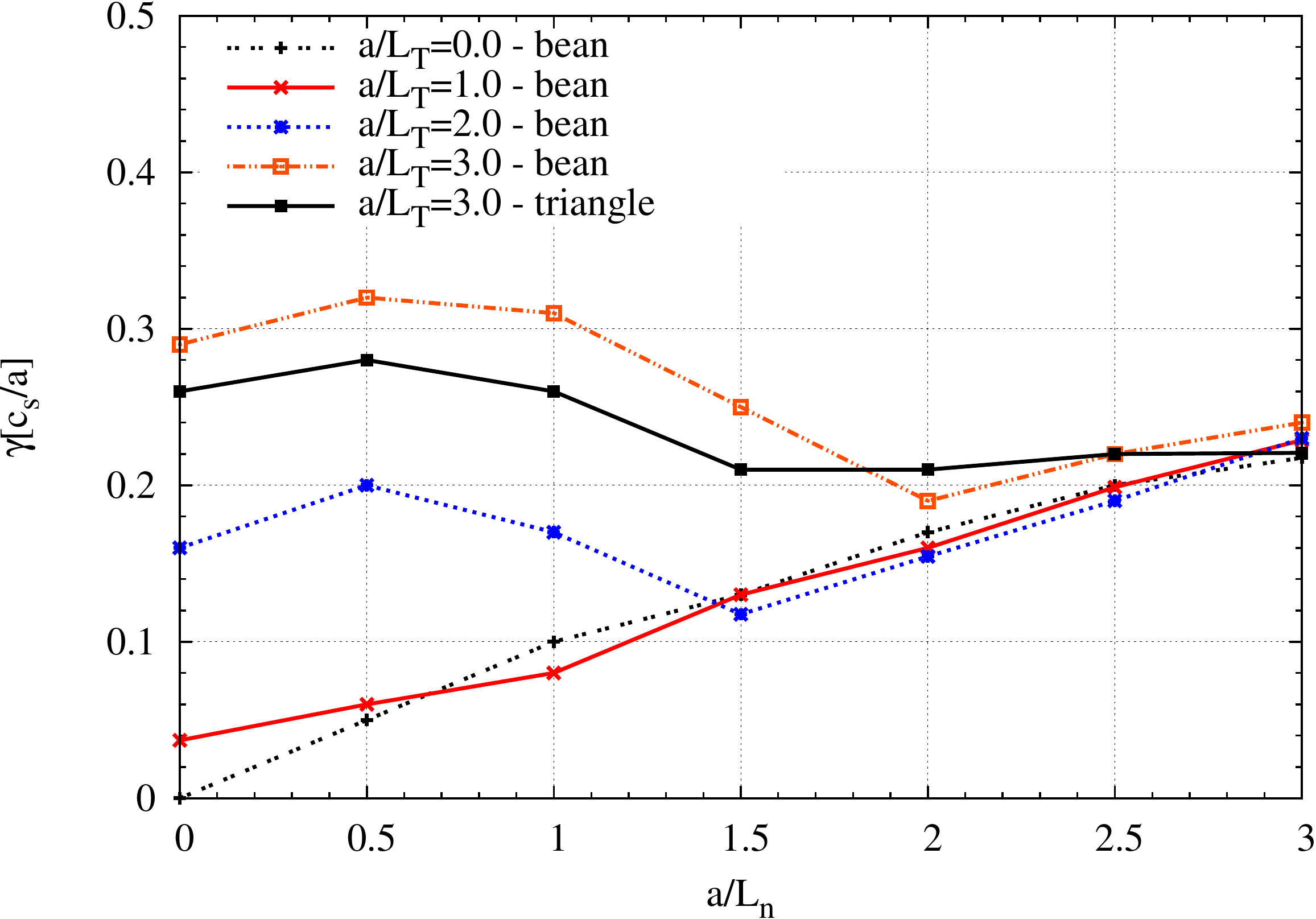}
\caption{\label{fig:ITGTEMW7X} Growth rates for mixed ITG-TEMs with $k_y\rho_s=1.1 $ in W7-X.}
\end{figure}

\begin{figure}
\includegraphics[width=0.45 \textwidth]{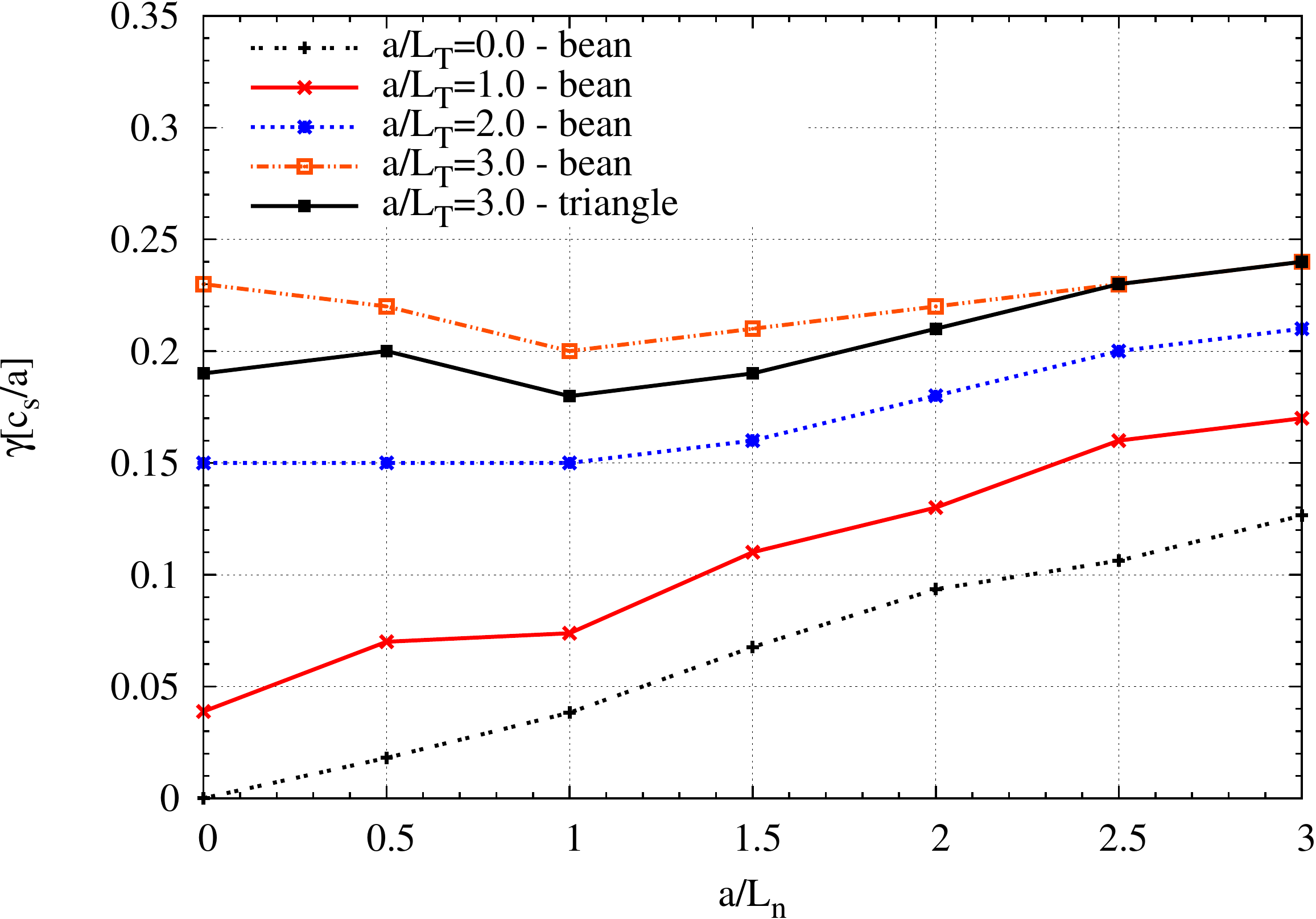}
\caption{\label{fig:ITGTEMQIPC} Growth rates for mixed ITG-TEMs with $k_y\rho_s=0.6 $ in QIPC.}
\end{figure}

In Wendelstein 7-X, it is not the TEM behaviour that survives but rather the ITG behaviour if both temperature gradients are switched on (Fig.\ \ref{fig:ITGTEMW7X}). Compared with the ITG simulations with kinetic electrons, the maximum growth rate is slightly elevated, but the general trends persist -- an ITG type mode for low density gradients and high temperature gradients, and trapped-particle modes for high density gradients. A similar mode transition was found in Ref.~ \cite{Xanthopoulos2007b} for ITG-TEM studies in W7-X. While the mode structures observed so far were usually either of ITG type, with their maxima located in the bad-curvature regions, or trapped-particle modes, peaking in the magnetic wells, here some truly mixed modes can be observed. They appear at $a/L_{T}=1.0$ and $a/L_{n}=0.5 \ldots 1.5$ and propagate in the ion diamagnetic direction. Concerning their mode structure, the perturbed electron temperatures exhibit trapped-particle features while the ion temperatures have peaks in the bad curvature regions. The resulting overall mode structure is thus a mixture of ITG and trapped-particle modes, but not a TEM since the electrons were again found to be stabilizing in an energy analysis for gradients of $a/L_{T}=1.0$ and $a/L_{n}=0.5$. The ratio of energy transfer was $\Delta E_e/\Delta E_i=-0.8$. The maximum growth rates were found at $k_y\rho_s=1.1 $. As before, the bean flux tube was identified as the more unstable one.\\
Also for QIPC, the growth rates found in the mixed ITG-TEM case resemble more those of the kinetic-electron ITG studies than those of the pure TEM simulations (Fig.\ \ref{fig:ITGTEMQIPC}). At high density gradients, the growth rates found here are much lower than in DIII-D, and the modes propagate in the ion direction. From the mode structure, it is evident that the trapped-particle mode is the dominant one. Only at vanishing density gradient are classical ITGs, destabilized by the temperature gradient, observed instead of the novel curvature-driven mode propagating in the electron diamagnetic direction found in the TEM case. As soon as there is a non-vanishing density gradient this becomes the dominant drive, even though the temperature gradient keeps exerting a destabilizing influence. Not surprisingly, the energy analysis at $a/L_{T}=1.0$ and $a/L_{n}=1.0$ reveals a ratio of energy transfer of $\Delta E_e/\Delta E_i=-0.6$, with the electrons being the stabilizing species. The most unstable mode was located at $k_y\rho_s=0.6 $, and the bean flux tube was the more unstable one.\\
\subsection{Comparison with analytical theory}
\begin{figure}
\includegraphics[width=0.45 \textwidth]{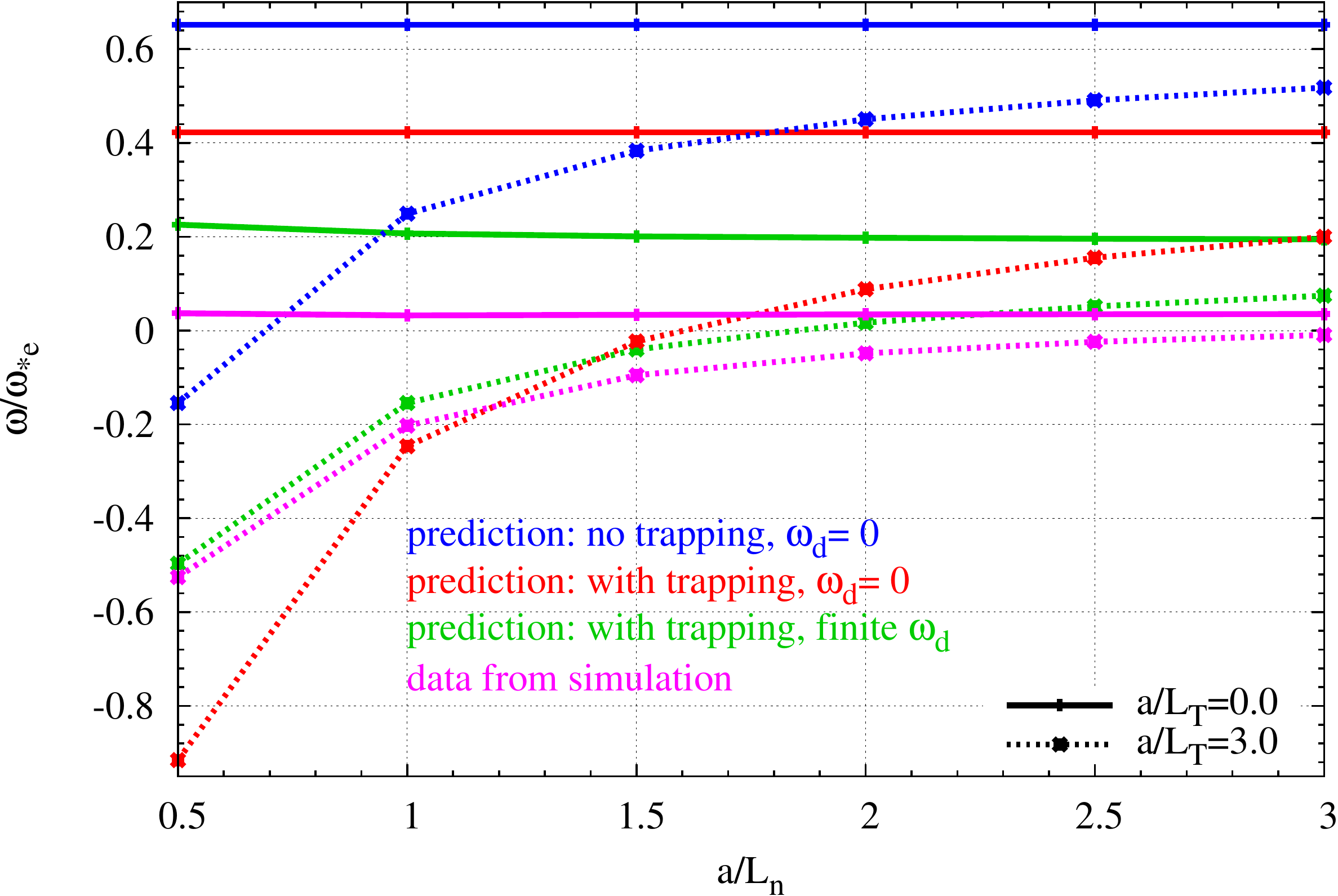}
\caption{\label{fig:eqncheckplotphitest} Comparison of the simulation results for $\omega/\omega_{*e}$ with the predicted values following Eqs. (\ref{eq:12}) and (\ref{eq:withtrapping}) for $k_y\rho_s=0.6$ in the ITG-TEM simulation of DIII-D using the analytical function for the potential}
\end{figure}

We now compare the simulation results of mixed ITG-TEMs obtained for DIII-D with the various analytical expressions that were found in Part I for the ratio between the real frequency of the mode, $\omega$, and the diamagnetic drift frequency of the electrons, $\omega_{*e}$. The simplest expression, neglecting the non-adiabatic response of the electrons and the finite magnetic drift frequency $\omega_d$ for both species is [Eq.\ (9) in Part I]
\bn
\frac{\omega}{\omega_{*e}}=\frac{\Gamma_0+\eta_i b \left(\Gamma_1-\Gamma_0\right)}{1+\frac{T_e}{T_i}\left(1-\Gamma_0\right)},
\label{eq:12}
\en
where $\Gamma_n=I_n(b)e^{-b}$ with $I_n$ the modified Bessel function and $b=k^2_{\bot}\rho_i^2$. Here $\omega/\omega_{*e}$ only depends on the ratio between ion temperature gradient and density gradient $\eta_i=L_{n}/L_{T_i}$ and on the wave number $k_y\rho_i=k_y\rho_s$ (due to $T_e=T_i$). In order to properly transition from $k_y\rho_s$ to  $k_{\bot}\rho_s=\sqrt{g^{yy}(l)k_y^2\rho_s^2}$, where $g^{yy}(l)$ is the metric element associated with the binormal direction that depends on the position $l$ along the field line, the average over the sinusoidal mode structure introduced in Part I was taken numerically, as suggested in Ref.~\cite{Dannert2005},
$$  
\frac{\omega}{\omega_{*e}}=\frac{\int \left(\Gamma_0+\eta_i b \left(\Gamma_1-\Gamma_0\right)\right)\phi^2 \frac{\mathrm{d}l}{B}}{\int \left(1+\frac{T_e}{T_i}\left(1-\Gamma_0\right)\right)\phi^2 \frac{\mathrm{d}l}{B}}.
$$
Equation (\ref{eq:12}) neglects any influence of the electron temperature gradient or any trapping effects and is thus not a very good prediction for the actually simulated values (see the blue curves in Fig.\ \ref{fig:eqncheckplotphitest}).
The first improvement to Eq.\ (\ref{eq:12}) is to include trapped electrons, which results in [Eq.\ (15) in Part I]
\begin{align}
& \frac{\omega}{\omega_{*e}}=\notag\\
& \frac{\int \left(\Gamma_0+\eta_i b \left(\Gamma_1-\Gamma_0\right)\right)\phi^2 \frac{\mathrm{d}l}{B}-\frac{1}{2}\int_{1/B_{max}}^{1/B_{min}}\Sigma_{\mathrm{wells}}\overline{\phi}^2(\lambda)\tau\mathrm{d}\lambda}{\int \left(1+\frac{T_e}{T_i}\left(1-\Gamma_0\right)\right)\phi^2 \frac{\mathrm{d}l}{B}-\frac{1}{2}\int_{1/B_{max}}^{1/B_{min}}\Sigma_{\mathrm{wells}}\overline{\phi}^2(\lambda)\tau\mathrm{d}\lambda}
\label{eq:withtrapping}
\end{align}
where the second part in both the numerator and the denominator accounts for the trapped particles. The first two terms in the numerator and denominator are equal to their counterparts in Eq.\ (\ref{eq:12}), except that they are now averaged along the field line and weighted by the squared amplitude of the mode structure. This integration is done numerically, again assuming a sinusoidal mode structure. Including the trapped particles in the analytical theory improves the result compared with the first estimate (see the red line in Fig.\ \ref{fig:eqncheckplotphitest} ), however including a finite magnetic drift frequency $\omega_d$ seems imperative.\\
This is done in the final step, where small but finite frequencies are allowed, $\omega_d \ll \omega $, using Eq. (16) of Part I, which results in a quadratic equation for $\omega/\omega_{*e}$. The resulting predictions come much closer to the actual results (see green line in Fig.\ \ref{fig:eqncheckplotphitest}). The deviation that can still be observed is most likely due to underestimating the magnitude of $\omega_d$. Comparing the real frequencies obtained in the simulations with the magnetic drift frequency it is found that $\omega_d / \omega = \mathcal{O}(1)$ and thus $\omega_d \ll \omega $ is not satisfied. This is particularly the case at vanishing temperature gradient, where $\omega$ is very small. In order to obtain yet better agreement, one might need to go to higher orders in the calculations. Also, the assumption of marginality is in general not fulfilled; the growth rate $\gamma$ was usually far from zero. In addition, the assumption of high frequencies $\omega \gg k_{\|}v_{Ti}$ was not met in most cases. These might be other reasons for the deviations between analytical theory and the simulation results. The choice of a sinusoidal function for the potential is, however, a good one, since the curves shown in Fig.\ \ref{fig:eqncheckplotphitest} come only marginally close to the measured values if the true, numerically found, function is used for the integration procedure. Our analytical theory is therefore well suited to predict the real frequencies of the modes.

\section{Conclusions and outlook}
\label{sec:conclusions}
In this paper, we have assessed the influence of the magnetic geometry, in particular quasi-isodynamicity, on the stability of ITGs and TEMs. It was found that, among the four configurations that were compared,  the two stellarators approaching quasi-isodynamicity and maximum-$J$ geometry, Wendelstein 7-X and QIPC, show significantly reduced growth rates compared with both the DIII-D tokamak and the quasi-axisymmetric stellarator NCSX. This stabilization seems indeed to occur thanks to the electrons drawing energy from the modes close to marginal stability. This effect is particularly important for TEMs, but the electrons were also found to be stabilizing for all configurations in case of ITGs. It could also be observed that QIPC, which is a stellarator that is more quasi-isodynamic than W7-X and whose regions of ``bad curvature'' are even smaller, is more stable than W7-X, which can be attributed to the enhanced quasi-isodynamicity. Moreover, none of the modes that were observed in W7-X and QIPC are classical electron-driven TEMs. The analytical prediction of quasi-isodynamic stellarators being resilient against classical TEMs, thus also holds for configurations that are only approximately quasi-isodynamic.\\

The next step should therefore be to simulate microinstabilities in finite-beta plasmas. 
All the simulations presented here were carried out in vacuum magnetic fields, but it is well known that quasi-isodynamicty and the maximum-$J$ property are much easier to achieve at high beta \cite{Rosenbluth1971b, Subbotin2006}. A finite plasma pressure should act stabilizing on the electrostatic instabilities we have studied, but could also excite electromagnetic modes, particularly as the ideal MHD stability limit is approached. Since the latter is fairly high in quasi-isodynamic stellarators, typically $\langle \beta \rangle \sim 5 - 10 \% $, there could be a region in parameter space where electrostatic instabilities are suppressed and electromagnetic ones not yet excited, resulting in a window of reduced turbulent transport. 

\begin{acknowledgments}
The authors thank F. Jenko, T. G\"orler, D. Told and T.M. Bird for many fruitful discussions as well as Yu. Turkin for providing the MCviewer for displaying the magnetic geometry.\\
Some of these simulations were performed on the HELIOS supercomputer, Japan.
\end{acknowledgments}

\end{document}